\documentclass[12pt]{article}
\usepackage[utf8]{inputenc}
\usepackage{amsthm}

\usepackage[colorlinks = true,
            linkcolor = blue,
            urlcolor  = blue,
            citecolor = blue,
            anchorcolor = blue]{hyperref}
  
\usepackage{titlesec}
\titleformat*{\section}{\large\bfseries} 
\titleformat*{\subsection}{\normalsize\bfseries}

\title{Bankruptcy Shocks and Legal Labor Markets:\\
Evidence from the Court Competition Era
}

\author{
\begin{tabular}
    {c@{\hskip 0.4in}c@{\hskip 0.4in}c}
    {Chad Brown\thanks{Department of Economics, University of Colorado Boulder, 80309, USA. chad.brown@colorado.edu}}
    & {Jeronimo Carballo\thanks{Department of Economics, University of Colorado Boulder, 80309, USA. jeronimo.carballo@colorado.edu}} 
    & {Alessandro Peri\thanks{Department of Economics, University of Colorado Boulder, 80309, USA. alessandro.peri@colorado.edu}} 
\\
\end{tabular}
}

\usepackage[authordate,bibencoding=auto,strict,backend=biber,natbib]{biblatex-chicago}
\usepackage{booktabs}   
\usepackage{lscape}
\usepackage{graphicx}   
\usepackage{amsmath}
\usepackage{amssymb}
\usepackage{float}      
\usepackage{enumerate}  
\usepackage{appendix} 
\usepackage{multicol} 
\usepackage[multiple]{footmisc} 
\usepackage[online,flushleft,para]{threeparttable} 
\usepackage{chngpage}
\usepackage[labelfont=bf]{caption} 
\usepackage{subcaption} 

\usepackage{adjustbox} 
\everypar{\looseness=-1}

\usepackage[margin=1.45in]{geometry} 

\usepackage{setspace}
\newtheorem*{definition*}{Definition}

\newcommand\mycaption[2]{\caption{\footnotesize\textbf{#1 -}  #2}}

\newcommand{\sym}[1]{\textsuperscript{#1}}

\newcommand{\tabnote}[8]{OLS regression estimates of #1 on: \textit{(i) Treatment:} #2; \textit{(ii) Fixed Effects:} #3; \textit{(iii) Controls:} #4. \textit{Sample:} #6. \textit{Period:} #7. \textit{Sources:} #8. 
    }

\newcommand{\colsepx}{\setlength{\tabcolsep}{13pt}}

\newcommand{\LP}{Lopucki BRD}
\newcommand{\LPlong}{UCLA-Lopucki Bankruptcy Research Database}
\newcommand{\Complong}{Standard and Poor's Compustat Database}

\newcommand{\Comp}{Compustat}
\newcommand{\CBP}{CBP}
\newcommand{\Census}{U.S. Census Bureau}
\newcommand{\BEA}{BEA}
\newcommand{\SEC}{SEC EDGAR}
\newcommand{\BLS}{BLS}
\newcommand{\PACER}{PACER}
\newcommand{\PCLlong}{PACER Case Locator}
\newcommand{\Bloomberg}{Bloomberg Law}

\newcommand{\lnLegEmp}{log Legal Employment}

\newcommand{\nBR}{number of bankruptcies in a county year}

\newcommand{\FS}{whether a firm filing for Ch11 bankruptcy chose to FS}

\newcommand{\sampleyears}{1991-1996}
\newcommand{\sampleyearsExtendedPeriod}{1991-2001}
\newcommand{\sampleyearsDROPNintyOne}{1992-1996}

\newcommand{\samplerestr}{All U.S. counties with non-zero legal employment, omitting Delaware and New York}
\newcommand{\samplerestrRBST}{all U.S. counties with non-zero legal employment, omitting Delaware and New York}
\newcommand{\sampleNYDE}{all U.S. counties with non-zero legal employment}
\newcommand{\sampleHQ}{subset of counties that had a publicly traded companies HQ at any point in the period 1991 through 1996, omitting Delaware and New York}
\newcommand{\sampleBRcounties}{all county-year observations that had atleast 1 bankruptcy, omitting Delaware and New York}

\newcommand{\sampleBRcourts}{subset of counties that where a bankruptcy court is located, omitting Delaware and New York}

\newcommand{\regressors}{number of bankruptcies, number of forum shopped bankruptcies, number of non-forum shopped bankruptcies}
\newcommand{\regressorsBRonly}{number of bankruptcies}
\newcommand{\regressorsDUMMY}{dummy variables indicating whether any bankruptcies or forum shopping occured}
\newcommand{\regressorsADJ}{number of bankruptcies, number of forum shopped bankruptcies, number of bankruptcies in adjacent counties}
\newcommand{\regressorsPLACEBO}{number of bankruptcies lagged one, two, or three years}
\newcommand{\regressorsPLACEBOdummy}{bankruptcy dummy lagged one, two, or three years}

\newcommand{\regressorsNYDESHOP}{number of bankruptcies, number of bankruptcies forum shopped to NY or DE}
\newcommand{\regressorsASSETS}{bankruptcy and forum shopping measured by the sum of assets for the firms at the time of their filing. Assets measured in billions of 2021 dollars}
\newcommand{\regressorsNoShock}{number of bankruptcies, number of forum shopped bankruptcies}

\newcommand{\controls}{log of county-year population, log of county-year employment in all non-legal sectors}
\newcommand{\controlsUNEMPRATE}{log of county-year population, log of county-year employment in all non-legal sectors, and county-year unemployment rate}
\newcommand{\controlsLAG}{log of previous years population for each county-year observation, log of previous years employment in all non-legal sectors for each county-year observation}

\newcommand{\controlsESTABCHANGE}{log of county-year population, log of county-year employment in all non-legal sectors, and county-year change in the log of the number of establishments in non-legal sectors from previous year}

\newcommand{\controlsCountyConditionChange}{county-year unemployment rate, and county-year change from previous year in the following-- log of population, log of employment in non-legal sectors, log of the number of establishments in non-legal sectors}

\newcommand{\source}{\LP, \Comp, \CBP, \Census, \BEA, \SEC}
\newcommand{\sourceAVGWAGE}{\LP, \Comp, \CBP, \Census, \BEA, \SEC, \BLS}

\newcommand{\countySY}{county and state-year}
\newcommand{\county}{county}
\newcommand{\countyyear}{county and year}
\newcommand{\countyDisY}{county and district-year, where district is one of 90 judicial districts}
\newcommand{\countyDivY}{county and division-year, where division is one of 9 Census Bureau-designated divisions}
\newcommand{\countyRegY}{county and region-year, where region is one of 4 Census Bureau-designated regions}
\newcommand{\FEall}{\county; \countyyear; \countyRegY; \countyDivY; \countySY; \countyDisY}

\newcommand{\clustercounty}{\textit{Standard Errors:} Clustered at county-level.}
\newcommand{\clusterfirm}{\textit{Standard Errors:} Clustered at firm-level.}

\newcommand{\sampleBRfirms}{County year observations of publicly traded firms who filed for Ch11 bankruptcy}
\newcommand{\controlsFSexog}{log of firm's total assets and employment, firm's leverage, log of county-year employment in legal sector, log of county-year average wage in legal sector and log of county-year population. Here we use $\text{Leverage}= \text{Liabilities}/\text{Assets}$}

\newcommand{\cjt}{c_{j,t}}
\newcommand{\njt}{n_{j,t}}
\newcommand{\wjt}{w_{j,t}}

\newcommand{\s}{n_{s}}

\renewcommand{\u}{n_{u}}

\newcommand{\wjct}{w^j_{c,t}}
\newcommand{\wsct}{w^s_{c,t}}
\newcommand{\wuct}{w^u_{c,t}}

\newcommand{\njct}{n^{j}_{c,t}}
\newcommand{\nsct}{n^{s}_{c,t}}
\newcommand{\nuct}{n^{u}_{c,t}}

\newcommand{\cjct}{c^{j}_{c,t}}

\newcommand{\eSct}{\epsilon_{c,t}^{nS}}
\newcommand{\eUct}{\epsilon_{c,t}^{nU}}

\newcommand{\NFSct}{\tilde c^{j}_{c,t}}
\newcommand{\NFSnt}{\tilde n^{j}_{c,t}}
\newcommand{\FSct}{\cjct}
\newcommand{\FSnt}{\njct}

\newcommand{\Zwi}{\mathbf{z}^{w}_{i}}
\newcommand{\Zni}{\mathbf{z}^{n}_{i}}
\newcommand{\Zi}{\mathbf{Z}_{i}}

\newcommand{\ewsi}{\epsilon_{i}^{wS}}
\newcommand{\ewui}{\epsilon_{i}^{wU}}
\newcommand{\eSi}{\epsilon_{i}^{nS}}
\newcommand{\eUi}{\epsilon_{i}^{nU}}
\newcommand{\ei}{\mathbf{u}_i}

\newcommand{\lc}{\boldsymbol{\lambda}_{c}}

\newcommand{\W}{\mathbf{W}}

\newcommand{\be}{\boldsymbol{\beta}}
\newcommand{\bgmm}{\hat{\be}_{\text{GMM}}}

\newcommand{\E}{\mathbb{E}}

\newcommand{\argmin}[1]{\underset{#1}{\text{argmin}}\:}

\newcommand{\sumin}{\sum_{i=1}^N}

\newcommand{\BR}{\text{\emph{BR}}}

\renewcommand{\footnoterule}{%
  \kern -3pt
  \hrule width 2in
  \kern 2pt
}
\makeatother

\begin{document}

\maketitle

\begin{abstract}
    \renewcommand{\thefootnote}{\fnsymbol{footnote}}
    
    \noindent We study how Chapter 11 bankruptcies affect local legal labor markets. 
    We document that bankruptcy shocks increase county legal employment and corroborate this finding by exploiting a stipulation of the law known as \emph{Forum Shopping} during the Court Competition Era (1991-1996). We quantify losses to local communities from firms forum shopping away from their local area as follows. First, we calculate the unrealized potential employment gains implied by our reduced-form results. Second, we structurally estimate a model of legal labor markets and quantify welfare losses. We uncover meaningful costs to local communities from lax bankruptcy venue laws.\\

    \end{abstract}

\pagebreak


\section{Introduction}
\vskip -0.8 em
\label{sec:Introduction}
    Legal expenses are often treated as dead-weight loss from bankruptcy proceedings (e.g.
\citealp{franks_comparison_1996};
\citealp{armour_costs_2012};
\citealp{m_m_buehlmaier_debt_2014};
\citealp{mahoney_bankruptcy_2015}).
However, they are meaningful transfers to crucial stakeholders of bankruptcy events: 
professionals employed in the legal service sector.
In this paper, we illustrate how failing to account for these transfers overlooks a potential source of economic stimulus, which acts in opposition to the overall contractionary effect 
that large bankruptcies have on local employment (\citealp{bernstein_bankruptcy_2019}). 
We provide first evidence that local legal labor markets indeed benefit from the increased demand for legal services created by bankruptcies. 
We then illustrate how lax bankruptcy venue laws and competition among courts may nullify this effect, by exporting away these potential gains from the affected communities. 
Finally, we develop a structural model of the legal sector and quantify economically meaningful welfare losses in the
bankrupt firm’s surrounding area from firms who file in courts far from their local communities. 

Little is known about the effect of corporate bankruptcies on local legal employment. 
For example, it could be negligible if  distressed firms hired legal services from a different locale, or if legal firms responded to the bankruptcy shock by partially increasing their employees' working hours or by reassigning them from other cases. 
Our first contribution is to document a positive and economically significant effect. 

Using a novel database of Chapter 11 bankruptcies of publicly traded firms,
we estimate that during the early 90's bankruptcy reorganizations are associated with a 1\% increase in annual legal employment for the affected county for each ongoing bankruptcy. 
Aggregated across the U.S. this means that each year roughly 10,000 jobs---with wages adding up to nearly \$1 billion in 2021 dollars---are being distributed to communities dealing with the contractionary effects associated with a major employer filing for bankruptcy.
Crucially, we document that these employment gains are lost when firms file for bankruptcy in courts that are far away from their headquarters, a phenomenon referred to as \textit{forum shopping}.
We find evidence that forum shopping \textit{exported away} more than 24\% of the potential employment gains from local communities with distressed firms each year during the Court Competition Era (1991-1996).
Building on the work of \citet{notowidigdo_incidence_2020}, we structurally estimate a labor model of the legal sector and document a $\approx$1\% consumption equivalent variation loss in affected communities as a result of forum shopping.\looseness=-1

Identifying the causal impact of bankruptcy shocks on local legal sector labor markets is not straightforward. Unlike the applied economics literature that exploits an event study framework, our research design setting is not conducive to such methods due to transitory and repeated treatment (\cite{callaway_difference--differences_2021}; \cite{sun_estimating_2020}). 
In order to address this problem, we need a shock that is: \textit{(i)} economically meaningful; \textit{(ii)} exogenous to the local demand for legal services; and \textit{(iii)} addresses the complex nature of our treatment.
We overcome this challenge by measuring demand shocks to local legal services using Chapter 11 reorganization bankruptcies of publicly traded firms. \looseness=-1

First of all, 
Chapter 11 bankruptcies act as a considerable demand shock to local legal services.
These bankruptcy cases are complex phenomena that arbitrate over the conflicting interests of several stakeholders: creditors, shareholders, workers, managers, and so forth. 
Consequently, firms will often pay fees in the order of hundreds of millions of dollars\footnote{
    For instance, the professional fees in Kodak bankruptcy case rounded up to \$240M in 2013. \textit{Source}: \href{https://www.reuters.com/article/us-kodak-bankruptcy/kodak-bankruptcy-advisers-likely-to-see-240-million-payday-idUSBRE9AD1DV20131114}{Reuters}\nocite{brown_kodak_2013}.  
    Professional fees averaged \$82 million in 2021 dollars in the \LPlong\, over the period 1980-2014.
    } 
for legal teams that range from a handful of highly trained consultants and attorneys, to teams of legal clerks and paralegals. 
%
Although it is common for firms to formulate a reorganization plan using a handful of bankruptcy experts from major hubs (like New York, \citealp{lopucki_courting_2005}), 
much of the impact of these shocks will be concentrated in the area surrounding whatever bankruptcy court the case was filed in. 
Lawyers representing the interested parties must be certified to practice in that area (\citealp{okray_attorney_2016}), familiar with local bankruptcy practices (\citealp{leary_standards_1999}), as well as 
able to attend to other in-person obligations
at the courthouse (e.g. bankruptcy filings, monitoring the docket, providing local counsel). 
Therefore, Chapter 11 bankruptcy shocks act as meaningful shocks to a local area. 


Second, Chapter 11 bankruptcy filings of publicly traded firms provide a plausible source of exogenous variation in the local demand for legal services.
The large/multi-national nature of the demand faced by publicly-traded firms 
decouples the determinants of firms' bankruptcy choices from the well being of the local economy.
To illustrate this point anecdotally, consider the Chapter 11 bankruptcy of Kodak, the well-known photography company from Rochester, New York. 
Economic fluctuations in Rochester played little to no role in Kodak's decision to file for bankruptcy reorganization in 2012---which was instead motivated by the worldwide decline in the popularity of film photography.
In Section \ref{sec:BR_LocalConditions}, 
we document this exogeneity empirically.

Third, we leverage
well-documented Chapter 11 forum shopping practices during the 90s to handle the complex nature of our treatment.
Counties experiencing treatment (Chapter 11 bankruptcy shocks) are often treated multiple times, can switch back and forth between treated and untreated from year to year, and do not share a clear pre-treatment and post-treatment period. 
%
We contribute to previous bankruptcy literature (e.g. \citealp{bernstein_bankruptcy_2019}; \citealp{chava2021community}) by proposing a novel identification strategy to handle the transitory and repeated nature of the treatment.
In place of an event study framework, 
we exploit forum shopping to create exogenous variations in the location of demand for legal service and verify our identification.
When a firm forum shops, the legal sector demand shock is effectively exported away from their local economy.
This transfer in potential legal service demand is particularly pronounced because firms typically forum shop considerable distances away from their local area. 
For large publicly traded corporations who forum shop the average distance from their headquarters to the court they filed in is 766.1 miles. 
This is a drastic increase from 18.9 miles for firms who did not forum shop.%
    \footnote{Figures are calculated using the \LPlong.}

To use forum shopping as an `institutional placebo', we follow an economic history approach and focus on a period where the determinants of firms' forum shopping decisions are clearly documented by the bankruptcy literature and arguably exogenous to local economic conditions:
the `Court Competition Era', 1991-1996 (Section \ref{sec:ForumShopping}).%
    \footnote{For example, \citet{cunningham_decriminalizing_2014} use a similar strategy to obtain identification. They use an institutional change where a Rhode Island judge unexpectedly decriminalized indoor prostitution to create exogenous variation in the availability and legality of sex services. As such they restrict their attention to the period from 2004 to 2009 when this institutional change was in effect.
    For other examples, see \citet{rajan_anatomy_2015}, \citet{hausman_fiscal_2016}. 
    }
From 1991 through 1996 bankruptcy judges began competing for prestigious and lucrative Chapter 11 filings (\citealp{EisenbergLopucki1999}),
which encouraged firm's to exploit stipulations of US bankruptcy law to file for bankruptcy far away from their headquarters, i.e. forum shop (e.g. \citealp{lopucki_courting_2005}; \citealp{Lopucki06}; \citealp{LopuckiKalin00}; \citealp{EisenbergLopucki1999}; \citealp{Cole03}). %
Over this short period of time, the share of forum shopped bankruptcy cases  tripled (Figure \ref{BR_FS_RATIO80_07}), from 20\% in 1990 to a new long run average of 60\%, which persists until today (\citealp{ellias_bankruptcy_2019}). %
%
Importantly, the bankruptcy literature suggests that forum shopping in this period is driven by incentives to seek favourable
judges
and is largely decoupled from local conditions 
(e.g. \citealp{EisenbergLopucki1999}; \citealp{lopucki_courting_2005}; \citealp{Lopucki06};  \citealp{gennaioli_judicial_2010}).%
    \footnote{
    See \citet{gennaioli_judicial_2010} for a study of the incentives structure of judges with 
    career concerns to compete for prestigious Chapter 11 reorganizations.
    } 
In Section \ref{sec:FS_LocalConditions}, we complement this literature and provide supportive evidence for the exogeneity of forum shopping with respect to unobserved factors correlated with attributes of local legal labor markets. 



\newcommand{\PACERlong}{Public Access to Court Electronic Records}

For our analysis,
we construct a novel database that expands the universe of large bankruptcies in the widely used \LPlong \space (Lopucki BRD) with bankruptcy information of publicly traded firms in Standard and Poor's Compustat Database (CS), 
supplemented with information using the \PACERlong\space (\PACER) service.%
        \footnote{
        Web BRD (the ‘‘LoPucki’’) database can be found at        \url{http://lopucki.law.ucla.edu}. 
        } 
After aggregating this information at the county level, 
we estimate a regression with county fixed effects, judicial district-year fixed effects, and county-year controls.\looseness=-1 

Our reduced-form analysis in Section \ref{sec:legempforsho} confirms that the observed changes in local legal employment following bankruptcy shocks are contingent on firms filing for bankruptcy in their local court. 
Thus, when firms file in a court that is far from their local area (i.e. forum shop) any impact on the legal sector is essentially exported away. 
This result indicates that it is the bankruptcy filing causing the measured effect and not confounding factors correlated with the bankruptcy.%
    \footnote{
    For example, we would still observe an increase in legal employment after forum shopping if the increase in legal employment was due to confounding factors associated with the bankruptcy event, such as spikes in divorces or wrongful termination lawsuits.}

Our reduced-form results are stable to a variety of robustness checks. Most notably, to ease concerns about the relevance of the comparison group we limit our sample to only counties that could plausibly be treated. 
We also run various specifications that show potential spillover effects to neighboring counties is negligible. 
Additionally we find our results to be robust to several other considerations such as: using alternative data sources or alternative measures of treatment, including additional county characteristic controls, and dropping years potentially impacted by recessions.

Besides providing a convenient `institutional placebo' to verify our identification, forum shopping practices have potentially important welfare consequences on the affected communities. 
To explore these effects, in Section \ref{sec:struct_est} we build on \citet{notowidigdo_incidence_2020} and construct a parsimonious model of the legal labor market.%
    \footnote{Our model is a specialization of the canonical labor model used for studying how shocks effect labor market outcomes across skill levels (\cite{acemoglu_technical_2002}; \cite{Acemoglu_Autor_2011}; \cite{bowlus_wages_2017}).}
In the model economy, a representative firm produces legal services using two inputs of production that act as imperfect substitutes: skilled labor (attorneys) and less productive unskilled labor (paralegals, law clerks, etc.).
In the spirit of \citet{bai_demand_2012}, bankruptcy demand shocks enter this model akin to ``productivity'' shocks to the production function of legal services, boosting the demand of legal workers and their welfare.

We calibrate the model parameters using a nonlinear simultaneous-equations generalized method of moments (GMM) procedure that exploits non-forum shopped bankruptcies as exogenous demand shocks for the local area.
Using this novel identification approach, we provide first empirical estimates for the elasticity of labor supply of skilled and unskilled workers in the legal industry.
This finding adds to the body of literature that estimates the Frisch elasticity of labor supply for the extensive margin (e.g. \cite{bianchi_icelands_2001}; \cite{brown_link_2013}; \cite{rogerson_micro_2009}). 
We provide insights into the labor markets of an industry that plays a key role in the functioning of modern economic societies (\citealp{bennett_corporate_2018}), but has been comparatively understudied with respect to similarly-sized sectors, like agriculture.%
    \footnote{In 2019 the legal industry GDP measured 280 billion, while the US agricultural sector measured 175 billion.
    \textit{Source:} \cite{FRED_Legal_Ag_GDP}.}


Our calibrated model allows us to to quantitatively estimate the local welfare implications of forum shopping. 
To do so we calculate the potential welfare gains associated with moving from the status-quo regime where firms forum shop to a counterfactual regime without forum shopping.
We find that legal sector workers in affected communities would need roughly a 1\% increase in there annual consumption to receive the same level of utility they would have had without forum shopping. \looseness=-1

These results 
additionally
contribute to the debate in the law literature over the efficiency of forum shopping and court competition for bankruptcy outcomes (e.g. \citealp{Zywicki06};  \citealp{ThomasRasmussen01}; \citealp{EisenbergLopucki1999}; \citealp{LopuckiKalin00}; \citealp{Lopucki06}; \citealp{lee_delawares_2011}), by providing first evidence of unintended consequences on the local economy. 
Accordingly, our findings have direct economic relevance on the ongoing policy debate in the U.S. Congress about the Bankruptcy Venue Reform Act.
On June 28, 2021 the U.S. Representatives Zoe Lofgren and Ken Buck introduced the bipartisan Bankruptcy Venue Reform Act to the U.S. House of Representatives.
If passed, the Reform Act will put a halt to the practice of forum shopping, the ability of firms to file for a bankruptcy court located in a different place than the principal assets or place of business.
The purpose of the bill is to protect the interests of affected stakeholders (local creditors, retirees and employees) by having the case processed by a judge familiar with the community%
    \footnote{See press release  \textit{Lofgren, Buck Introduce Bipartisan Legislation to End Corporate Bankruptcy Forum Shopping}, from Rep. Lofgren. 
    \textit{Source:} \href{https://lofgren.house.gov/media/press-releases/lofgren-buck-introduce-bipartisan-legislation-end-corporate-bankruptcy-forum}{Press Release}.
    }
and put to an end ``the single most significant source of injustice in Chapter 11 bankruptcy cases''.%
    \footnote{City of Detroit’s Bankruptcy Judge Steven Rhodes, \textit{National Conference of Bankruptcy Judges}, 2018. \textit{Source:} \href{https://cdn.ymaws.com/www.ncbj.org/resource/resmgr/docs_public/Venue_White_Paper_-_Final.pdf}{NCBJ Special Committee}.}
Our analysis suggests that a deeper look at the production network of large corporate bankruptcy cases uncovers economically significant costs to local communities, so far unaccounted for by the policy debate.

\section{Forum Shopping and the Court Competition Era}
\vskip -0.8 em
\label{sec:ForumShopping}
    The U.S. federal courts system is organized in 90 judicial districts which handle bankruptcy cases.%
        \footnote{In our analysis, we do not include the district court for Puerto Rico or the territorial courts for Guam, the Northern Mariana Islands, or the Virgin Islands.}
    Pursuant to \href{https://www.justice.gov/jm/civil-resource-manual-189-bankruptcy-jurisdiction-venue}{28 U.S.C. \textsection1408}, distressed firms can file for Chapter 11 reorganization in any judicial district 
    with jurisdiction over 
    the location of their (1) ``principal place of business''; (2) ``principal assets''; (3) ``domicile''; or in which (4) the debtor’s affiliate, general partner, or partnership has a pending bankruptcy case.
    The practice of filing in a venue which is located in a district different from the principal place of business or principal assets is referred to as forum shopping. 
    %
    Consistent with the previous bankruptcy literature,
    we follow the procedure adopted by the \LP\, and consider
    the firms' headquarters to be the locale of the principal place of business or principal assets.
    
\begin{figure}[h]
        \centering
        \includegraphics[width =.65\textwidth]{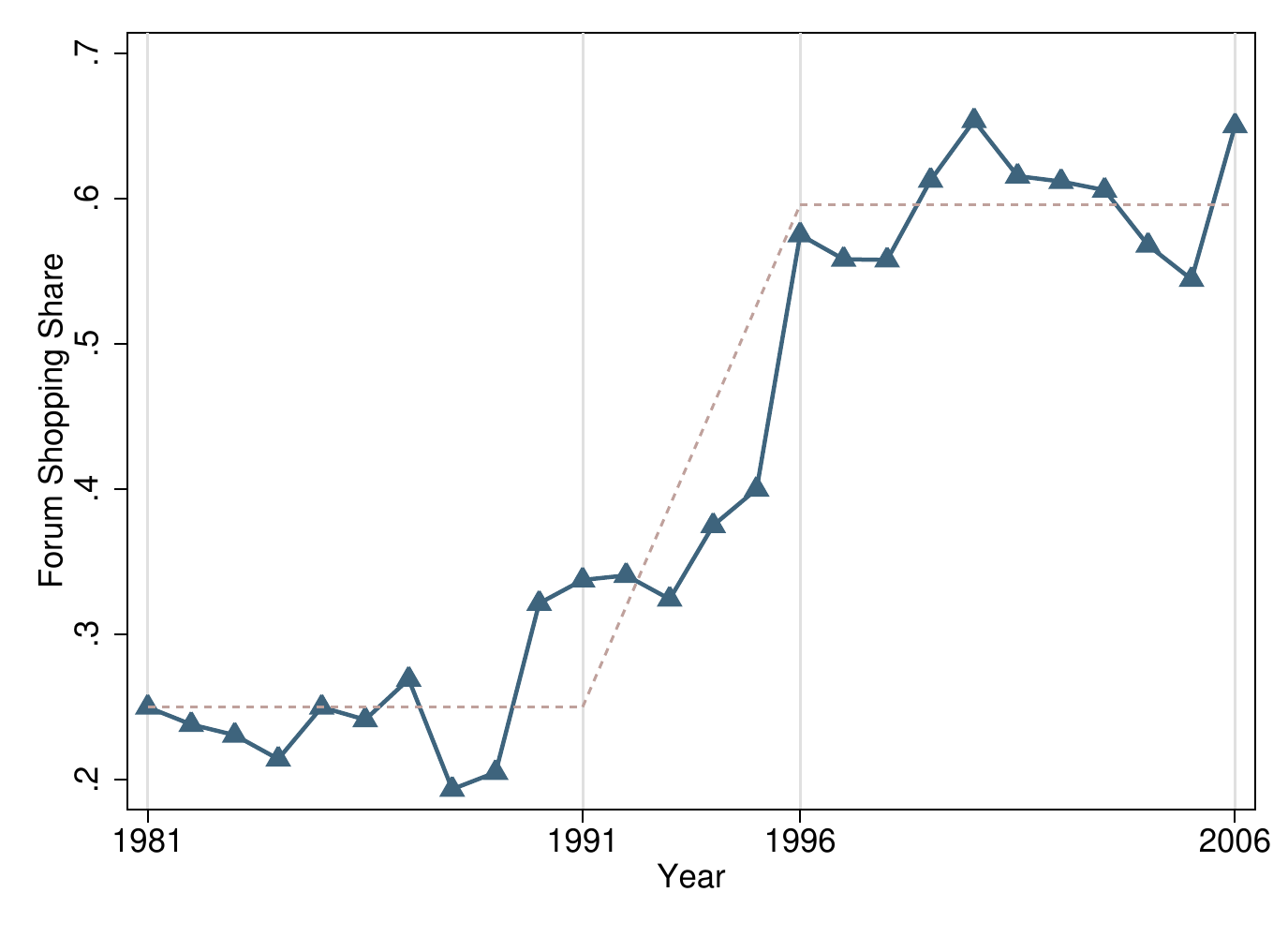}
        \mycaption{Share of Forum Shopped Bankruptcies}{Share of the Chapter 11 Bankruptcies that were Forum Shopped annually from 1981 to 2006.
        Dashed lines show the average share of forum shopped Ch11 bankruptcies before and after the court competition era, and the change in the average over the court competition era.
        Due to limited data availability for the extended period, the figure only uses data from the \LP \space for large firm bankruptcies.}
        \label{BR_FS_RATIO80_07}
    \end{figure}

    The period from 1991 through 1996 witnessed an explosion in forum shopped filings across the U.S. 
    In the brief time window of six years, the share of forum shopped cases nearly tripled, leveling off at roughly 60\% in 1997 from an average of 25\% prior to 1991 (Figure \ref{BR_FS_RATIO80_07}).
    Many bankruptcy scholars attribute this boom in forum shopping to an intense competition between the Southern District of New York and the District of Delaware, driven by judges who wished to attract prestigious and lucrative large Chapter 11 filings\footnote{
        \citet{Cole03} finds that ``[a]lmost all of the judges suggested that there is a level of prestige and satisfaction that attaches to hearing and deciding important cases....'Big Chapter 11 cases are interesting as well as prestigious.'' For a model of judicial discretion in corporate bankruptcy, see \cite{gennaioli_judicial_2010}.} 
     (e.g. \citealp{LopuckiKalin00}; \citealp{Lopucki06}; \citealp{lopucki_courting_2005}; \citealp{gennaioli_judicial_2010}).
     
\begin{figure}[h]
        \centering
        \includegraphics[width =.65\textwidth]{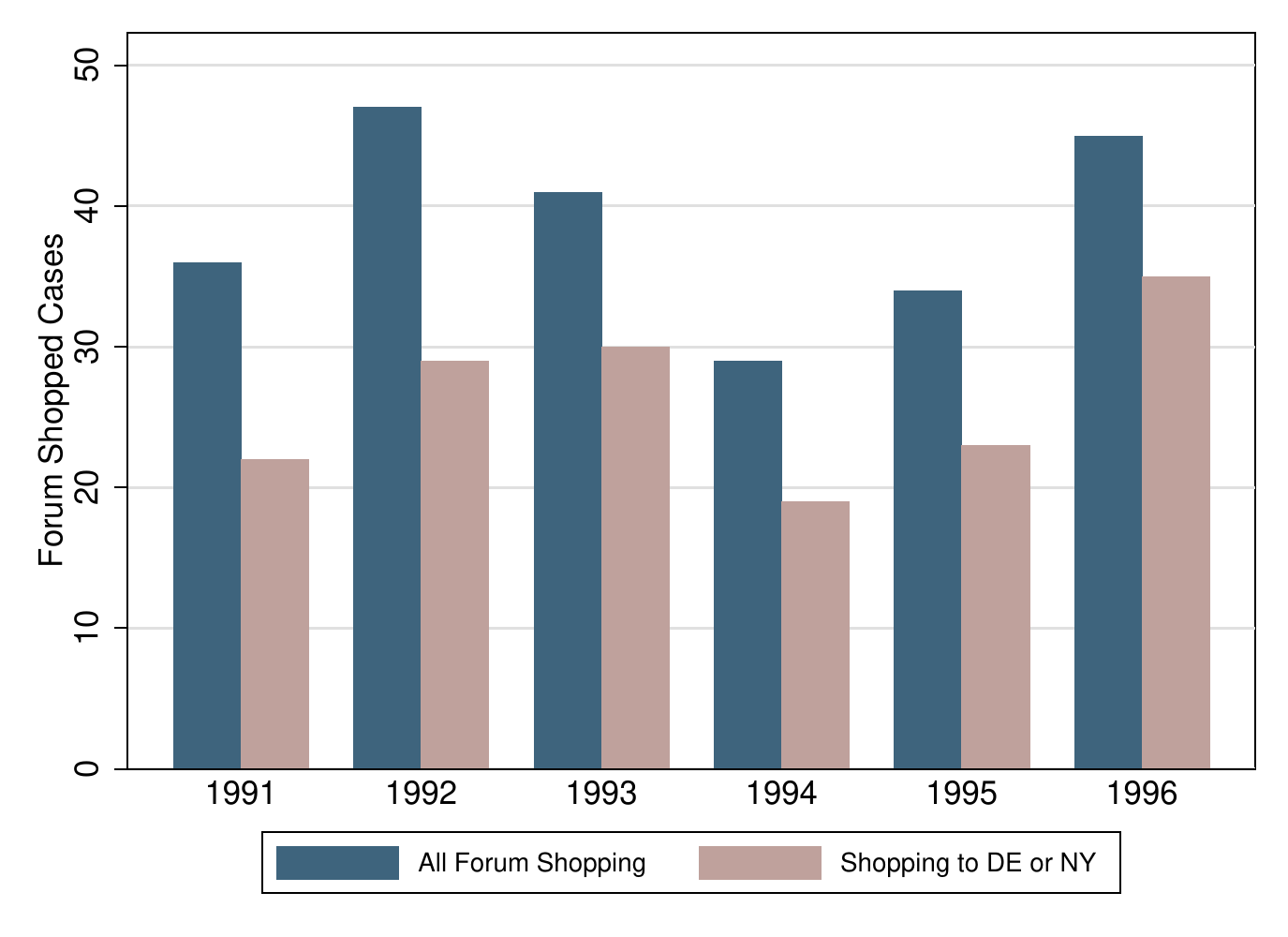}
        \mycaption{Forum Shopped Bankruptcies}{Number of Forum Shopped Ch11 Bankruptcies filing in Delaware or New York Southern vs total number of Forum Shopped Ch11 Bankruptcies.}
        \label{FS_share}
    \end{figure}

    While New York was the favored destination for forum shopping prior to the 1990's, this abruptly changed in the early 90's when Delaware emerged as a prominent forum shopping destination.
    Delaware's meteoric rise in popularity marked the beginning of the `Court Competition Era'. 
    Firms started exploiting the flexible venue criteria to `shop around'\footnote{
        \citet{lopucki_venue_1991} lists several examples of companies ``venue hook", where firms select a venue by quickly transferring headquarters to a small office near their preferred court shortly before filing. In addition after a 1989 ruling firms were also able to exploit flexible criteria for place of incorporation to forum shop.} 
    for a venue where they believed judges would be more likely to rule in their favor.
    Figure \ref{FS_share} illustrates the extent of New York and Delawares prominence during this time.
    
    This trend of forum shopping and court competition continued through 1996.
    Then controversy over this trend caused the National Bankruptcy Review Commission to propose a reform that would have essentially eliminated debtors from filing in Delaware (\citealp{skeel_1998} p. 1-2).
    In 1997, to avoid this impending legislation, the Chief Judge of the Delaware District Court made an order to reduce the number of filings from forum shopping debtors as well as reduce the ability for judges to compete for prestigious bankruptcy cases (\citealp{skeel_1998, LopuckiKalin00}).%
        \footnote{\citet{skeel_1998} (p. 33-35), also offers other explanations but stresses the importance of the proposal for the Judge's order.}$^,$%
        \footnote{See \citet{LopuckiKalin00} p. 234-- Delaware Senator Joseph Biden controlled bankruptcy legislation in the Senate and is speculated to have not put the law to a vote to allow Delaware to deal with the problem on their own terms and allow forum shopping to continue in a reduced form.
        }
    So, while Delaware remains a hotbed for forum shopping even today, the reduction in court competition among judges has made the motives and outcomes of forum shopping less clear (\citealp{lee_delawares_2011}).%

\section{Empirical Analysis}\label{sec:Empirical_Methodology}
\vskip -0.8 em
    This section lays out the empirical strategy for estimating the impact of Chapter 11 bankruptcies on local legal markets. 
    Section \ref{sec:data} describes the construction of our Chapter 11 bankruptcy database. 
    Section \ref{sec:sample} describes our main estimation sample and Section \ref{sec:DescStat} shows descriptive statistics for bankruptcy treatment during the sample.
    Section \ref{sec:Methodology} presents our empirical methodology.
    Finally, Section \ref{sec:IdentificationChallenges} tests 
    that Chapter 11 bankruptcies in our sample are exogenous to local conditions, and that forum shopping decisions are unrelated to local legal market conditions. 
    Henceforth, bankruptcy refers only to Chapter 11 cases.

    \subsection{Data}\label{sec:data}
\vskip -0.8 em
        Our main database covers the period 1991-1996 and includes county level data for all U.S. states.
        This amounts to 18,373 county-year observations. 
        The database includes annual-county data from the County Business Pattern (CPB) database. For each county, it also includes detailed data on bankruptcies and forum shopping for firms headquartered in the county. 
        The bankruptcy information combines  firm-level bankruptcy information from the \LPlong\; (\LP) and \Complong\; (\Comp) with public court records. 
        In the next subsection, we explain how we construct our main database. 
        Additional details can be found in Appendix \ref{App:Data_Creation}.

        \subsubsection{Bankruptcy Data}
\vskip -0.5 em
        Our bankruptcy database integrates the \LP\, with bankruptcy information from Compustat and public court records over the period 1991-1996. 
        
        The \LP\; contains information on 182 Chapter 11 bankruptcy cases of \textit{large} publicly listed US companies during 1991-1996. 
        Firms qualify as large public companies only if they file a 10-K form with the SEC that reports assets worth at least \$100 million in 1980 dollars (about \$300 million in 2021 dollars).
        
        We complement this database with bankruptcy information of smaller publicly listed firms from \Comp.
           This database contains panel data on 13,277 publicly traded US firms over 1991-1996. 
        We follow the procedure in \citet{corbae_reorganization_2021} 
        to determine Chapter 11 bankruptcy events in our sample (see Appendix \ref{App:Data_Creation}).
        %
        We then turn to publicly available SEC
        filings and Court records to
        identify the bankruptcy cases that have been forum shopped.
            \footnote{
            \textit{Sources:} \href{https://www.sec.gov/edgar/search-and-access}{SEC.gov | EDGAR - Search and Access}, and \href{https://pcl.uscourts.gov}{PACER Case Locator}, supplemented by \href{https://www.bloomberglaw.com}{bloomberglaw.com} as needed.  \nocite{SEC_EDGAR}
            } 
        We proceed in two steps.
        First, we use these records
        to determine the location of the court of filing for each bankruptcy. 
        Second, we establish that a bankruptcy case has been forum shopped if the bankruptcy was filed in a different judicial district than where the firm's headquarter is located.
        By doing so, we integrate the \LP\, with additional 
        385 bankruptcy cases, of which 18\% were forum shopped.

        \subsubsection{Local Economic Attribute Data}
\vskip -0.5 em
        We use County Business Patterns (CBP) data from the US Census Bureau to measure the county-year level of employment (as imputed by \citealp{CBP}) in the legal sector for our dependent variable, and to construct different controls (such as employment in non-legal sectors) for our empirical analysis.\footnote{
            \textit{Source:} \href{http://fpeckert.me/cbp/}{Imputed CBP Data}-- \textcite{CBP}.}
         CBP provides administrative data on employment and establishment counts for more than 900 industries for each county and year for the period we focus (see \citealp{CBP}). The granularity of the industry classification available in the CBP allows us to focus on the legal sector employment (SIC 8100) and avoid measurement issues such as the ones that could come from aggregation with other services.
         Average wages for both skilled and unskilled legal workers are from the Bureau of Labor Statistics (BLS).\footnote{
             \textit{Source}: \href{https://www.bls.gov/data/}{Bureau of Labor Statistics Databases}. \nocite{BLS_Database}}  
        County population data comes from United States Census Bureau's intercensal data.\footnote{\textit{Source:} \href{https://www.census.gov/en.html}{U.S. Census Bureau- Intercensal Datasets}.\nocite{census}
            }

    \subsection{Sample}\label{sec:sample}
\vskip -0.8 em
        We follow the forum shopping literature (e.g., \cite{LopuckiKalin00}), and focus our attention to the Court Competition Era, the period from 1991 to 1996 when Delaware emerged as a forum shopping destination, effectively creating a duopoly competition with New York (Section \ref{sec:ForumShopping}). 
        Our sample consists of all 
        U.S. counties with non-zero legal employment over the years 1991-1996. 
        To avoid contamination effects for counties that \textit{receive} forum shopped cases, our main sample omits counties from New York (NY) and Delaware (DE).
        This eases concerns that our results may be affected by counties receiving forum shopping cases, because these two states receive nearly all forum shopped cases (see Figure \ref{FS_share}).%
            \footnote{ 
            Additionally we perform robustness checks to confirm that that our main results are not driven by this omission (see Section \ref{sec:robust_check}).}

    \subsection{Bankruptcy Descriptive Statistics}\label{sec:DescStat}
\vskip -0.8 em
        
        Table \ref{tab:DescriptiveStats_OnlyBR_FS} reports descriptive statistics for bankruptcies and forum shopping in our sample.%
            \footnote{
            See Table \ref{tab:Extra_Desc_Stats} in Appendix \ref{App:SumStat} for an additional descriptive table that includes the controls and the dependent variable used in our main specification.}
        Our sample consists of approximately 544 bankruptcies, where roughly one-quarter of the cases are forum shopped. 
        For the 451 county-year observations with at least one bankruptcy, we record on average 1.70 bankruptcy fillings and 0.47 forum shopped cases. 
        
        \begin{table}[H]
            \centering
            \begin{threeparttable}[]
                \caption{Descriptive Statistics}
                {
\def\sym#1{\ifmmode^{#1}\else\(^{#1}\)\fi}
\begin{tabular}{l*{2}{cc}}
\toprule
                    &\multicolumn{2}{c}{\textbf{Full Sample}}&\multicolumn{2}{c}{\textbf{County-Year with BR}}\\
                    &        Mean&   Std. Dev.&        Mean&   Std. Dev.\\
\midrule
Bankruptcies        &       0.048&       0.397&       1.703&       1.673\\
\addlinespace
Forum Shopping      &       0.013&       0.155&       0.470&       0.800\\
\midrule
Observations        &       16064&            &         451&            \\
\bottomrule
\end{tabular}
}
 
                \label{tab:DescriptiveStats_OnlyBR_FS}
                \begin{tablenotes}
                    \footnotesize\textit{Note:} County level descriptive statistics for number of bankruptcies and forum shopped bankruptcies. \textit{Sources:} \source.\\
                    \item \textbf{Full Sample}: All counties with non-zero legal employment during the period 1991-1996, not including counties from New York and Delaware.\\         
                    \item \textbf{County-Year with BR}: All counties where at least one bankruptcy took place during the period 1991-1996, not including counties from New York and Delaware. 
                \end{tablenotes}        
            \end{threeparttable}
        \end{table}

        Figure \ref{BR_FS_Industry} illustrates the distribution of forum shopping across SIC industry groups. The figure shows that both bankruptcy and forum shopping took places in virtually all industries. Similarly,  Appendix \ref{App:Figures} shows that the geographical distribution of bankruptcies (Figure \ref{BR_County_Map}) and forum shopping (Figure \ref{FS_County_Map}) is fairly well spread out across the U.S.
        with clusters of bankruptcies not surprisingly occurring in counties with major population centers. 
        
        

        
        \setcounter{figure}{4}
        \begin{figure}[h]
            \begin{subfigure}[b]{0.5\textwidth}
                \centering
                \includegraphics[width = \textwidth]{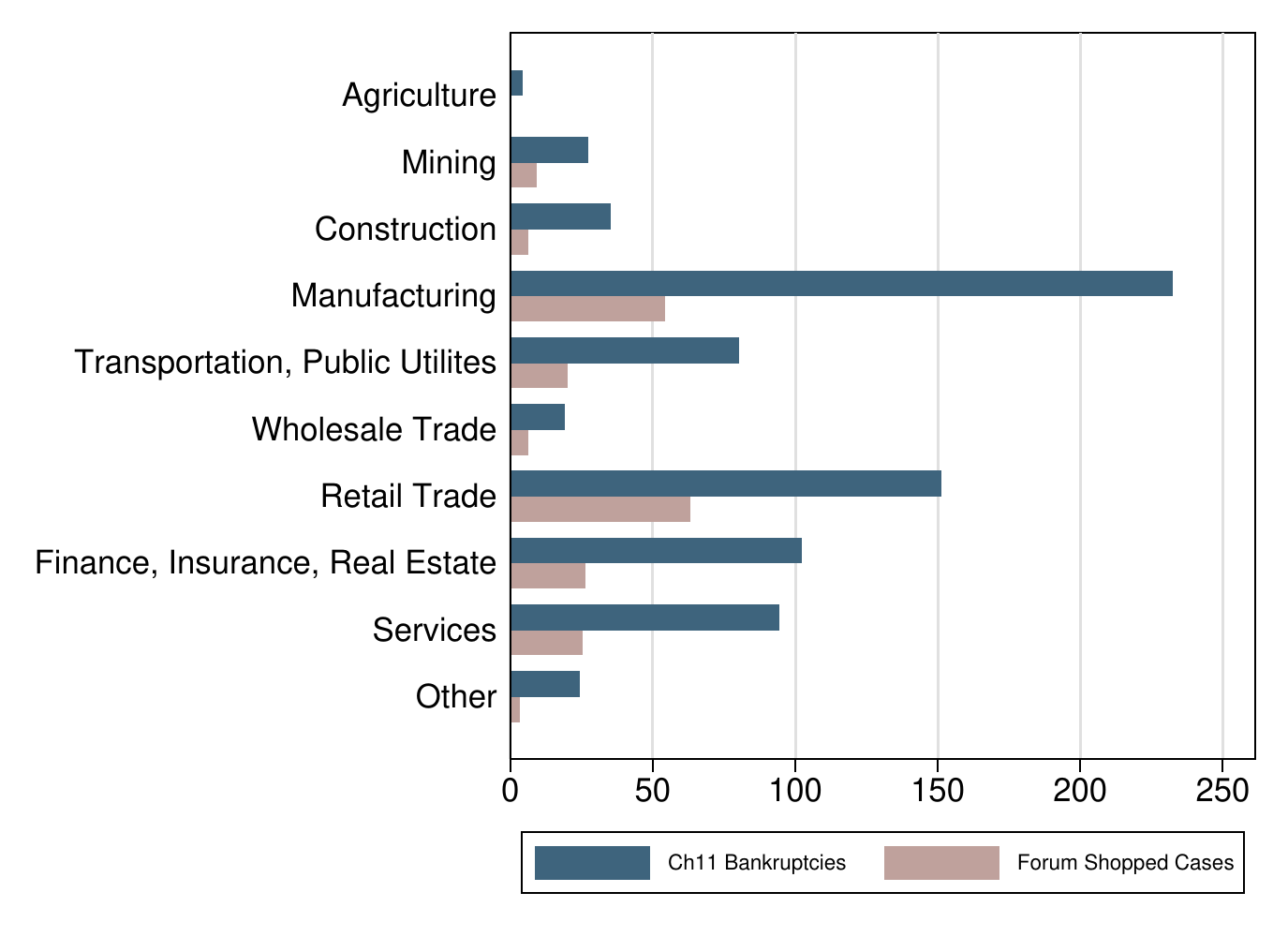}
                \label{BR_FS_Industry1}
            \end{subfigure}
            \begin{subfigure}[b]{0.5\textwidth}
                \centering
                \includegraphics[width = \textwidth]{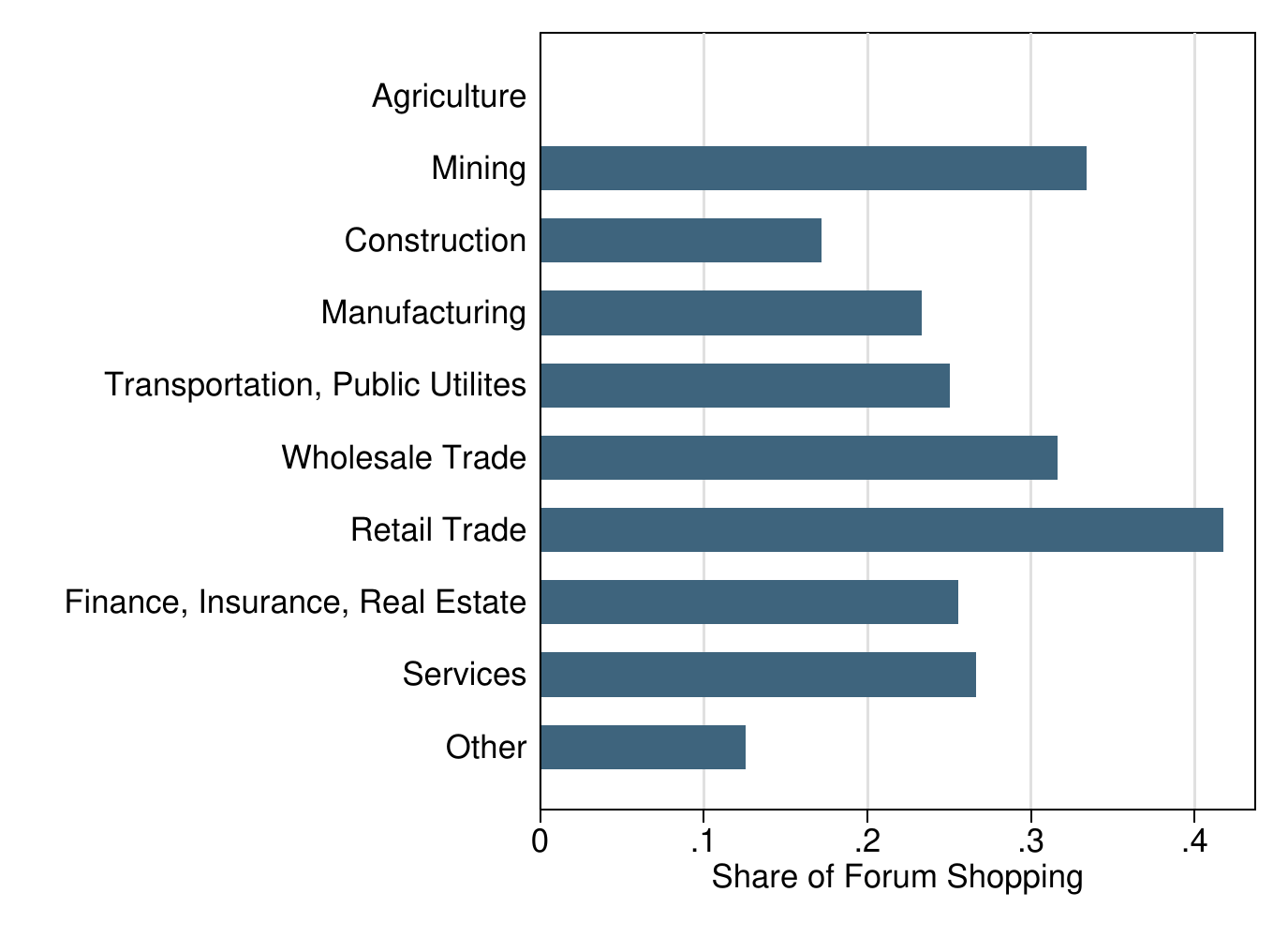}
                \label{BR_FS_Industry2}
            \end{subfigure}
            \mycaption{Bankruptcy and Forum Shopping by Industry}{Number of Ch11 bankruptcies and forum shopped Ch11 Bankruptcies (left); share of Ch11 bankruptcies that are forum shopped by industry (right). Both graphs show cases from 1991 through 1996, omitting cases from New York and Delaware. Industry is defined by SIC division.}\label{BR_FS_Industry}
        \end{figure}

    \subsection{Methodology}\label{sec:Methodology}
\vskip -0.8 em
    
        To study the impact of Chapter 11 bankruptcy filings on the local legal sector, we estimate a linear regression model,
        \begin{equation}
           \ln y_{ct} = \gamma_{BR} \times \text{\emph{bankruptcies}}_{ct} + \beta X_{ct}  + \eta_c + \delta_{d(c)t} + \varepsilon_{ct}\;.
           \label{eq:spec_bench_BR}
        \end{equation} 
        $y_{ct}$ captures our main outcome of interest: the level of employment in the legal sector in county $c$ in year $t$.
        The treatment variable $\emph{bankruptcies}$ denotes the county-year bankruptcy shock. 
        As our benchmark measure, we use the number of firms headquartered in county $c$ that have an active Chapter 11 filing in year $t$. 
        This measure includes forum shopped cases.
        
        The parameter $\gamma_{BR}$ is our key estimation target. 
        We expect $\gamma_{BR} >0$ if bankruptcies generate a positive demand shock to local legal markets. As mentioned in the introduction, this is far from a foregone conclusion. It could be the case that bankruptcy filings do not increase local employment due to legal firms reassigning workers from other cases, hiring workers in other counties, or increasing the hours worked without hiring more workers, among other things.
        
        To estimate $\gamma_{BR}$, we exploit identifying variation coming from Chapter 11 bankruptcies of large publicly traded firms.
        The bankruptcy decisions of these large corporations---usually multinational firms---are largely independent from fluctuations in the local economy (county) where they are headquartered, which absorb only a small fraction of their profits and revenues. 
        As a result, these bankruptcies provide plausibly exogenous shocks to the local legal labor market. Section \ref{sec:BR_LocalConditions} corroborates this exogeneity assumption.
        
        We account for other potential sources of variation in the level of legal employment with time-varying controls at the county level and fixed effects. 
        The matrix $X_{ct}$ captures the additional time-varying controls, such as the log of county's population and the log of county's employment level in non-legal sectors in a given year. 
        We expect both of these to be positively correlated with the level of legal sector employment.
        The county level fixed effects, $\eta_{c}$, captures county-specific time-invariant factors that affect the level of legal employment. 
        District-year fixed effects, $\delta_{d(c)t}$, absorbs all time-varying variables at the district-year level that affect our outcome of interest. 
        These time-varying factors include measures of the business cycle at the district level, as well as potential policy changes for particular district courts. 
        After dropping counties from New York and Delaware, our sample is left with 85 judicial districts that subdivide the remaining states into between 1 and 4 parts.
        Accordingly, district-year fixed effects control for time-varying unobserved factors at the state level.
        %
        Finally, $\varepsilon_{ct}$ is the usual error term.

        To exploit forum shopping as an `institutional placebo',
        we estimate the following augmented model
        \begin{equation}
           \ln y_{ct} = \gamma_{BR} \times \text{\emph{bankruptcies}}_{ct} + \gamma_{FS} \times \text{\emph{forum shopping}}_{ct} + \beta X_{ct}  + \eta_c + \delta_{d(c)t} + \varepsilon_{ct}\;.
           \label{eq:spec_bench_BR_FS}
        \end{equation} 
        where $\gamma_{FS}$ captures the effect of forum shopping on our main outcome. 
        For firms headquartered in county $c$ during year $t$,
        $\emph{forum shopping}_{ct}$ 
        counts the number of firms included in $\emph{bankruptcies}_{ct}$ 
        that have filed in a district that is different than the district where the firm's headquarters are located.%
            \footnote{
            To identify the local effect of bankruptcy, we need a clean geographical separation between the bankrupt's area 
            and the court of filing.
            To do so, we define a forum-shopped case, as a bankruptcy case filed in a different district than the one where the firm is headquartered. The BRD database also considers in-district' forum shopping, which only account for a handful of cases (11 out 88).
            %
            } 
        Note that by construction $0 \le \text{\emph{forum shopping}}_{ct} \le \text{\emph{bankruptcies}}_{ct}$. 
        
        This specification has two objectives.
        The first goal is to  test our identification strategy by verifying  that $\gamma_{BR}$ effectively measures the impact of bankruptcies on local legal employment. If corporate bankruptcies are merely acting as a proxy for another determinant of demand for legal services (e.g., worker lawsuits, a spike in divorces, etc.), then the location of the filing should not affect our estimated results. However, if the estimated effect of forum shopping is negative and `cancels out' the effect of the bankruptcy, then it seems plausible that our observed effect is  a result of the bankruptcy filing itself.  This first goal requires that the decision to forum shop is not systematically related to attributes of the local legal markets. 
        For example, if firms forum shop due to high wages in the local legal sector, then forum shopping would not provide the required exogenous variation to work as an `institutional placebo'. 
        In Section \ref{sec:FS_LocalConditions}, we provide institutional background and relevant literature to argue why we do not expect this to be the case.
        In addition, we perform a regression analysis to corroborate this exogeneity assumption.
        
        The second goal is to 
        improve the measurement of the effect of bankruptcy filings on local legal employment, by allowing regular and forum shopped filings to have a differential impact. 
        

    \subsection{Identification Challenges}\label{sec:IdentificationChallenges}
\vskip -0.8 em
    In this section, we address potential concerns over two key identifying assumptions: \textit{(i)} exogeneity of bankruptcy shocks to changes in local county conditions;  \textit{(ii)} exogeneity of firms' choice to forum shop to local legal market conditions. 
    

        \subsubsection{Bankruptcy Shocks Exogenous to Local Conditions}\label{sec:BR_LocalConditions}
\vskip -0.5 em
        Publicly traded firms are large (often multinational) firms with customers across the U.S. and around the world. Accordingly, the share of profits that directly depends on the local conditions of their headquarter county is very small and
        is unlikely to have any impact on their
        decision to file for bankruptcy. Just as economic fluctuations in Rochester, New York played little to no role in Kodak's decision to file for bankruptcy, bankruptcy decisions of other publicly traded firms in our sample are plausibly decoupled from their local economy. 
        They 
        are beholden to fluctuations in demand on a much larger scale than their local county.
        As a result, Chapter 11 bankruptcies of publicly traded firms can plausibly provide 
        identifying variation to estimate the effect of demand shocks on local legal markets services.
        
        
        \begin{table}[h]
            \adjustbox{max width=\textwidth, center = \textwidth}{%
            \begin{threeparttable}
            \caption{Regressions for number of bankruptcies on changes in county conditions.}
            
            {
\def\sym#1{\ifmmode^{#1}\else\(^{#1}\)\fi}
\begin{tabular}{l*{5}{c}}
\toprule
                    &\multicolumn{1}{c}{(1)}         &\multicolumn{1}{c}{(2)}         &\multicolumn{1}{c}{(3)}         &\multicolumn{1}{c}{(4)}         &\multicolumn{1}{c}{(5)}         \\
\midrule
\%$\Delta$(County Population)&       0.221         &       0.224         &       0.273         &       0.233         &       0.288         \\
                    &     (0.209)         &     (0.210)         &     (0.204)         &     (0.212)         &     (0.207)         \\
\addlinespace
\%$\Delta$(Emp Non-Legal)&                     &    -0.00402         &                     &                     &    -0.00356         \\
                    &                     &   (0.00880)         &                     &                     &   (0.00927)         \\
\addlinespace
\%$\Delta$(Non-Legal Establishments)&                     &                     &     0.00206         &                     &     0.00601         \\
                    &                     &                     &    (0.0188)         &                     &    (0.0200)         \\
\addlinespace
Unemployment Rate   &                     &                     &                     &     0.00134         &     0.00144         \\
                    &                     &                     &                     &   (0.00171)         &   (0.00172)         \\
\midrule
County FE           &         Yes         &         Yes         &         Yes         &         Yes         &         Yes         \\
District-Year FE    &         Yes         &         Yes         &         Yes         &         Yes         &         Yes         \\
R-Squared           &       0.772         &       0.772         &       0.774         &       0.772         &       0.774         \\
Observations        &       15931         &       15931         &       15925         &       15931         &       15925         \\
\bottomrule
\multicolumn{6}{l}{\footnotesize Standard errors clustered at county level in parentheses}\\
\multicolumn{6}{l}{\footnotesize \sym{*} \(p<0.1\), \sym{**} \(p<0.05\), \sym{***} \(p<0.01\)}\\
\end{tabular}
}

            \label{tab:BR_LocalConditions}
            
            \begin{tablenotes}
                \item [] \footnotesize\textit{Note:} \tabnote{\nBR}{NA}{\countyDisY}{\controlsCountyConditionChange}{\clustercounty}{\samplerestr}{\sampleyears}{\sourceAVGWAGE}
            \end{tablenotes}
            \end{threeparttable}
            }
        \end{table}

            To corroborate this exogeneity assumption, we empirically evaluate whether bankruptcy filings of publicly traded firms are indeed uncorrelated with fluctuations in their headquarter county's socio-economic conditions.
            To do so, we regress the number of county-year bankruptcies on several county-year economic attributes that can be correlated with local county conditions, such as unemployment rate at the county level, changes in county population, entry and exit of firms at the county level, and changes in employment level outside of the legal sector.
            Table \ref{tab:BR_LocalConditions} shows evidence that changes in county population, employment, number of establishments in non-legal sectors, and even unemployment 
            do not predict bankruptcy filings. 
            These results suggest that any observed changes in local legal employment when a county experiences a bankruptcy shock is unlikely to have been caused by confounding factors related to changes in county conditions.\looseness=-1   
            
            

            This plausible exogeneity to confounding factors is further bolstered by a battery of
            placebo tests (Section \ref{sec:Placebo}), robustness checks (Section \ref{sec:robust_check}), and by exploiting forum shopping as an `institutional placebo' (Section \ref{sec:legempforsho}).
        \subsubsection{Forum Shopping Exogenous to Local Conditions}\label{sec:FS_LocalConditions}
\vskip -0.5 em

        To exploit forum shopping for identification, 
        we need firms choice to forum shop to be conditionally exogenous with respect to county attributes which affect legal sector employment.
        We support this claim in two ways. 
        First, we refer to the relevant literature to illustrate how this exogeneity assumption has institutional roots in Chapter 11 bankruptcy practices.
        Then, we empirically evaluate whether firms' decisions to forum shop are correlated with local legal labor market attributes, conditional on firms' size, degree of financial distress and fixed effects. \looseness=-1
        
        Much of the literature relating to forum shopping during the Court Competition Era contends that the primary incentive for firms to forum shop was to find ``debtor-friendly'' courts (e.g. \citealp{lopucki_courting_2005}; 
        \citealp{gennaioli_judicial_2010}). 
        For instance, \citet{EisenbergLopucki1999} gives evidence against alternative incentives for forum shopping and ascertains that ``the persistence of forum shopping demonstrates the importance of judges to litigants''. 
        

         Furthermore, the exogeneity of the firm's forum shopping decision is consistent with their modus operandi for hiring bankruptcy experts.
        When a large publicly traded firm files for Chapter 11 bankruptcy, regardless of where they file, it is typical to hire a handful of legal experts from large hubs such as New York to develop a reorganization plan (\citealp{lopucki_courting_2005}). 
        Therefore, it is unlikely that firms 
        chose to forum shop because the legal industry in their area lacked sufficient expertise.
        The pattern of forum shopping during the Court Competition Era
        suggests that the quality of the legal industry in the filing court's area was not a major factor in the firm's venue choice.
        Indeed, when Delaware emerged as a hot bed for forum shopping in the early 1990's, it was considered to be a ``sleepy, backwater bankruptcy district virtually devoid of bankruptcy professionals'' (\citealp{EisenbergLopucki1999}).
        
         \begin{table}[h]
            \adjustbox{max width=\textwidth, center = \textwidth}{%
            \begin{threeparttable}
            \caption{Regressions for firms' choice to forum shop on firm and county characteristics.}
            
            {
\def\sym#1{\ifmmode^{#1}\else\(^{#1}\)\fi}
\begin{tabular}{l*{5}{c}}
\toprule
                    &\multicolumn{1}{c}{(1)}         &\multicolumn{1}{c}{(2)}         &\multicolumn{1}{c}{(3)}         &\multicolumn{1}{c}{(4)}         &\multicolumn{1}{c}{(5)}         \\
\midrule
ln(Firm Assets)     &      0.0642\sym{***}&                     &      0.0642\sym{***}&      0.0748\sym{***}&      0.0798\sym{***}\\
                    &    (0.0189)         &                     &    (0.0195)         &    (0.0217)         &    (0.0289)         \\
\addlinespace
Firm Leverage       &     0.00179         &                     &     0.00214\sym{*}  &     0.00265\sym{*}  &   -0.000544         \\
                    &   (0.00115)         &                     &   (0.00124)         &   (0.00135)         &   (0.00334)         \\
\addlinespace
ln(Firm Employment) &     0.00496         &                     &     0.00614         &    -0.00831         &     -0.0182         \\
                    &    (0.0174)         &                     &    (0.0179)         &    (0.0199)         &    (0.0307)         \\
\addlinespace
ln(County Legal Sector Emp)&                     &     -0.0163         &     -0.0419         &     -0.0361         &      -0.111         \\
                    &                     &    (0.0495)         &    (0.0541)         &    (0.0539)         &     (0.104)         \\
\addlinespace
ln(County Legal Sector Avg Wage)&                     &    -0.00843         &       0.100         &       0.133         &       0.344         \\
                    &                     &     (0.144)         &     (0.155)         &     (0.157)         &     (0.447)         \\
\addlinespace
ln(County Population)&                     &      0.0340         &      0.0430         &      0.0251         &      0.0524         \\
                    &                     &    (0.0544)         &    (0.0579)         &    (0.0586)         &     (0.116)         \\
\midrule
Fixed Effects       &                     &                     &                     &                     &                     \\
Year                &         Yes         &         Yes         &         Yes         &         Yes         &          No         \\
Industry            &          No         &          No         &          No         &         Yes         &         Yes         \\
District-Year       &          No         &          No         &          No         &          No         &         Yes         \\
\midrule
R-Squared           &       0.112         &      0.0231         &       0.117         &       0.143         &       0.464         \\
Observations        &         351         &         393         &         344         &         343         &         251         \\
\bottomrule
\multicolumn{6}{l}{\footnotesize Standard errors clustered at firm level in parentheses}\\
\multicolumn{6}{l}{\footnotesize \sym{*} \(p<0.1\), \sym{**} \(p<0.05\), \sym{***} \(p<0.01\)}\\
\end{tabular}
}

            \label{tab:FS_EXOG}
            
            \begin{tablenotes}
                \item [] \footnotesize\textit{Note:} \tabnote{\FS}{NA}{year, industry, and district-year}{\controlsFSexog}{\clusterfirm}{\sampleBRfirms}{\sampleyears}{\sourceAVGWAGE}
            \end{tablenotes}
            \end{threeparttable}
            }
        \end{table}


       To provide empirical support for these  institutional arguments, we regress an indicator of whether or not a firm forum shopped on firms' characteristics and local legal sector labor markets characteristics in the year of filing. Since only firms that filed for bankruptcy can decide to forum shop, we restrict our sample to bankrupt firms. 
       To measure firms' size we use total assets and total employment. We then include leverage to control for the firms' degree of financial distress.
        We use legal sector employment, average wage in the legal sector and county population as county characteristics. 
        
        Table \ref{tab:FS_EXOG} presents the results of these regressions with a variety of fixed effects.
        Across all specifications, we find no evidence that a firm's propensity to forum shop is influenced by any of the local legal sector characteristics.
        Column (2) shows this finding holds even when nothing is controlled for besides county characteristics and year fixed effects.  
        We find that only the total assets of the firm at the time of bankruptcy are associated with a meaningfully significant increase in the likelihood of a firm to forum shop, indicating that firm size may play a role. \looseness=-1

\section{Bankruptcy Shocks and Legal Labor Markets}\label{sec:Results}
\vskip -0.8 em
    This section estimates the effect of bankruptcy shocks on local legal labor markets.
    A large publicly listed firm bankruptcy is a significant demand shock to legal services. Average professional fees for one of these bankruptcies amount to \$82 million in 2021 dollars in our Lopucki BRD sample.
    %
    While some of these fees may go to bankruptcy experts located in hubs like New York (\citealp{lopucki_courting_2005}), much of the fees go to legal services performed where the bankruptcy court is located.\looseness=-1
    These services are carried in locus by teams of legal clerks and paralegals certified to practice in the bankruptcy court area, familiar with local bankruptcy practices, and able to attend proceedings and process paperwork at the courthouse. \citet{freedman_how_2005} estimates that it takes upwards of 1.3 non-lawyer employees for every lawyer. 
    


    Section \ref{sec:legemp} estimates the impact of bankruptcies on county legal employment. 
    Section \ref{sec:legempforsho} illustrates how results change when we consider forum shopping. 
    Section \ref{sec:robust_check} demonstrates the robustness of our results to additional considerations.
    Finally Section \ref{sec:Pot_Emp_Gain} builds on the previous results to estimate the potential employment and wage gains lost to New York and Delaware because of forum shopping.


    \subsection{Bankruptcy Shocks and Legal Employment}\label{sec:legemp}
\vskip -0.5 em
        We use our benchmark model \eqref{eq:spec_bench_BR} to estimate the effect of bankruptcy shocks on legal employment. 
        Table \ref{tab:MAIN_90_97_BRonly} reports estimates
        obtained by regressing the county-year level of legal employment on the number of bankruptcies and county-year controls.
        These results indicate that a bankruptcy filing from a firm headquartered in the county is associated with a statistically and economically significant increase in the level of legal employment in that locale. 
        From left to right, each column shows the estimates for gradually more demanding fixed effects along the geographical and time dimension.\footnote{
            We follow the Census Bureau definition of the region and use 4 categories: Northeast, Midwest, South and West. In the case of Column (6), we introduce judicial district-year fixed effects. 
            The U.S. states are divided into 90 judicial districts that subdivide the states into 1-4 parts. Hence, this is akin to controlling for time-varying within-state factors.
            }
        The impact of one bankruptcy on legal employment is robust and stable across the different specifications, varying between 0.070\% and 0.086\%.
        
        
        \begin{table}[h]
            \adjustbox{max width=\textwidth, center = \textwidth}{%
            \begin{threeparttable}
            \caption{Employment Level Regressions}
            
            {
\def\sym#1{\ifmmode^{#1}\else\(^{#1}\)\fi}
\begin{tabular}{l*{6}{c}}
\toprule
                    &\multicolumn{1}{c}{(1)}         &\multicolumn{1}{c}{(2)}         &\multicolumn{1}{c}{(3)}         &\multicolumn{1}{c}{(4)}         &\multicolumn{1}{c}{(5)}         &\multicolumn{1}{c}{(6)}         \\
\midrule
Bankruptcies        &     0.00793\sym{***}&     0.00830\sym{***}&     0.00720\sym{**} &     0.00768\sym{**} &     0.00702\sym{**} &     0.00859\sym{***}\\
                    &   (0.00280)         &   (0.00287)         &   (0.00330)         &   (0.00335)         &   (0.00283)         &   (0.00284)         \\
\addlinespace
ln(County Population)&       0.797\sym{***}&       0.750\sym{***}&       0.726\sym{***}&       0.715\sym{***}&       0.651\sym{***}&       0.694\sym{***}\\
                    &     (0.115)         &     (0.125)         &     (0.137)         &     (0.144)         &     (0.134)         &     (0.134)         \\
\addlinespace
ln(Emp Non-Legal)   &       0.181\sym{***}&       0.166\sym{***}&       0.167\sym{***}&       0.170\sym{***}&       0.174\sym{***}&       0.170\sym{***}\\
                    &    (0.0418)         &    (0.0468)         &    (0.0481)         &    (0.0489)         &    (0.0479)         &    (0.0474)         \\
\midrule 
Fixed Effects       &                     &                     &                     &                     &                     &                     \\
County              &         Yes         &         Yes         &         Yes         &         Yes         &         Yes         &         Yes         \\
Year                &          No         &         Yes         &          No         &          No         &          No         &          No         \\
Region-Year         &          No         &          No         &         Yes         &          No         &          No         &          No         \\
Division-Year       &          No         &          No         &          No         &         Yes         &          No         &          No         \\
State-Year          &          No         &          No         &          No         &          No         &         Yes         &          No         \\
District-Year       &          No         &          No         &          No         &          No         &          No         &         Yes         \\
\midrule   
R-Squared           &       0.984         &       0.984         &       0.984         &       0.984         &       0.984         &       0.984         \\
Observations        &       15963         &       15963         &       15963         &       15931         &       15963         &       15963         \\
\bottomrule
\multicolumn{7}{l}{\footnotesize Standard errors clustered at county level in parentheses}\\
\multicolumn{7}{l}{\footnotesize \sym{*} \(p<0.1\), \sym{**} \(p<0.05\), \sym{***} \(p<0.01\)}\\
\end{tabular}
}

            \label{tab:MAIN_90_97_BRonly}
            
            \begin{tablenotes}
                \item [] \footnotesize\textit{Note:} \tabnote{\lnLegEmp}{\regressorsBRonly}{\FEall}{\controls}{\clustercounty}{\samplerestr}{\sampleyears}{\source}
            \end{tablenotes}
            \end{threeparttable}
            }
        \end{table}

        Our preferred specification is the most demanding and controls for county and district-year fixed effects (Table \ref{tab:MAIN_90_97_BRonly} Column (6)).
        These fixed effects absorb all determinants of legal employment level that are constant at the county level and absorb all time-varying determinants at the district level, including district GDP, unemployment rate, as well as any other district level measure of the business condition. 
        Under this specification, 
        each active bankruptcy filing is associated with a 0.77\% increase in county legal employment. 
        Hence, using the figures from Table \ref{tab:DescriptiveStats_OnlyBR_FS}, counties with at least one ongoing bankruptcy are associated with a 1.3\% increase in legal employment. Using \BLS\; data on average wages these job losses come out to roughly \$5 million worth of employment annually for the affected county. 
   

        \subsubsection{Placebo Tests}\label{sec:Placebo}
            We perform several placebo tests to evaluate whether there were differences in the local legal employment level in the counties affected by bankruptcies before the filings took place. To do this, we lag the variable that captures the number of bankruptcies. More specifically, we use three alternative lag structures: one, two, and three years before bankruptcies actually occurred. Additionally, we combine all three measures in the same specification. 
            
            {\setlength{\tabcolsep}{12pt}        
            \begin{table}[h]
            \adjustbox{max width=\textwidth, center = \textwidth}{%
            \begin{threeparttable}
            \caption{Employment Level Regressions with lags for bankruptcy shock.}
            
            {
\def\sym#1{\ifmmode^{#1}\else\(^{#1}\)\fi}
\begin{tabular}{l*{4}{c}}
\toprule
                    &\multicolumn{1}{c}{(1)}         &\multicolumn{1}{c}{(2)}         &\multicolumn{1}{c}{(3)}         &\multicolumn{1}{c}{(4)}         \\
\midrule
L1.Bankruptcies     &     0.00231         &                     &                     &     0.00256         \\
                    &   (0.00312)         &                     &                     &   (0.00312)         \\
\addlinespace
L2.Bankruptcies     &                     &    0.000304         &                     &   -0.000529         \\
                    &                     &   (0.00280)         &                     &   (0.00256)         \\
\addlinespace
L3.Bankruptcies     &                     &                     &   -0.000118         &    0.000549         \\
                    &                     &                     &   (0.00315)         &   (0.00335)         \\
\addlinespace
ln(County Population)&       0.715\sym{***}&       0.716\sym{***}&       0.715\sym{***}&       0.715\sym{***}\\
                    &     (0.144)         &     (0.144)         &     (0.144)         &     (0.144)         \\
\addlinespace
ln(Emp Non-Legal)   &       0.170\sym{***}&       0.170\sym{***}&       0.171\sym{***}&       0.171\sym{***}\\
                    &    (0.0489)         &    (0.0489)         &    (0.0490)         &    (0.0490)         \\
\midrule
County FE           &         Yes         &         Yes         &         Yes         &         Yes         \\
District-Year FE    &         Yes         &         Yes         &         Yes         &         Yes         \\
R-Squared           &       0.984         &       0.984         &       0.984         &       0.984         \\
Observations        &       15931         &       15931         &       15930         &       15930         \\
\bottomrule
\multicolumn{5}{l}{\footnotesize Standard errors clustered at county level in parentheses}\\
\multicolumn{5}{l}{\footnotesize \sym{*} \(p<0.1\), \sym{**} \(p<0.05\), \sym{***} \(p<0.01\)}\\
\end{tabular}
}

            \label{tab:placebo}
            
            \begin{tablenotes}
                \item [] \footnotesize\textit{Note:} \tabnote{\lnLegEmp}{\regressorsPLACEBO}{\countyDisY}{\controls}{\clustercounty}{\samplerestr}{\sampleyears}{\source}
            \end{tablenotes}
            \end{threeparttable}
            }
        \end{table}
            }
            
            Table \ref{tab:placebo} shows that none of the lagged bankruptcy measures are associated with a significant change in legal employment
            and the estimated coefficients shrink by an order of magnitude compared to the contemporaneous version. 
            Hence, there are no significant differences in the level of local legal employment for the counties affected by bankruptcies before the bankruptcies actually occur.  
            These results provide further evidence for our claim that bankruptcy filings are associated with a positive impact on the local legal employment
            level. See Appendix \ref{App:Placebo} for additional placebo tests using alternative treatment measures.

    \subsection{Forum Shopping}\label{sec:legempforsho}
\vskip -0.5 em
        From 1991 through 1996, the intense competition between Delaware and New York ignited a surge in forum shopping cases all around the US  (see Section \ref{sec:ForumShopping}), as filing firms flocked to these two magnet districts.
        For these forum shopping firms, the distance between where the firms are located and where they file is considerably larger. 
        In the \LP\; sample, the average distance from firm headquarters to the court of filing for forum shopped cases was 766.1 miles from the firms headquarters, compared to just 18.9 miles for non-forum shopped cases. 
        As a result, 
        these bankruptcy cases should not trigger the usual spike in demand for legal services in the firms local area, and therefore should not affect the level of local legal employment.
        
        {\setlength{\tabcolsep}{16pt}
        \begin{table}[h]
            \adjustbox{max width=\textwidth, center = \textwidth}{%
            \begin{threeparttable}
            \caption{Employment Level Regressions}
            
            {
\def\sym#1{\ifmmode^{#1}\else\(^{#1}\)\fi}
\begin{tabular}{l*{3}{c}}
\toprule
                    &\multicolumn{1}{c}{(1)}         &\multicolumn{1}{c}{(2)}         &\multicolumn{1}{c}{(3)}         \\
\midrule
Bankruptcies        &     0.00768\sym{**} &      0.0100\sym{***}&                     \\
                    &   (0.00335)         &   (0.00373)         &                     \\
\addlinespace
Non-FS Bankruptcies &                     &                     &      0.0100\sym{***}\\
                    &                     &                     &   (0.00373)         \\
\addlinespace
Forum Shopping      &                     &     -0.0111\sym{*}  &    -0.00108         \\
                    &                     &   (0.00637)         &   (0.00564)         \\
\midrule
County FE           &         Yes         &         Yes         &         Yes         \\
District-Year FE    &         Yes         &         Yes         &         Yes         \\
R-Squared           &       0.984         &       0.984         &       0.984         \\
Observations        &       15931         &       15931         &       15931         \\
\bottomrule
\multicolumn{4}{l}{\footnotesize Standard errors clustered at county level in parentheses}\\
\multicolumn{4}{l}{\footnotesize \sym{*} \(p<0.1\), \sym{**} \(p<0.05\), \sym{***} \(p<0.01\)}\\
\end{tabular}
}

            \label{tab:MainEmpLvl}
            
            \begin{tablenotes}
                \item [] \footnotesize\textit{Note:} \tabnote{\lnLegEmp}{\regressors}{\countyDisY}{\controls}{\clustercounty}{\samplerestr}{\sampleyears}{\source}
            \end{tablenotes}
            \end{threeparttable}
            }
        \end{table}
        }

        To verify this conjecture Table \ref{tab:MainEmpLvl} uses model specification \eqref{eq:spec_bench_BR_FS}, which controls for both bankruptcy and forum shopped filings.\footnote{
            We also estimate alternative versions of Table \ref{tab:MainEmpLvl} where we use different sets of fixed effects.
            These tables are reported in Appendix \ref{App:Fixed_Effects} and in all cases they confirm our results.} 
        Column (2) shows that for a county-year observation, the estimated effect of the number of bankruptcies on the legal 
        sector employment is 1\%, a notable increase from the previous specification that did not control for forum shopping (0.7\%). 
        In addition, forum shopping has a negative impact of a similar magnitude, dampening the effect of bankruptcies not filed locally. 
        %
        %
        Column (3) tests the net effect of a forum shopped bankruptcy filing on local legal employment by using an alternative measure of bankruptcy that only includes bankruptcies filed locally (non-forum shopped). 
        As expected, this shows that a forum shopped bankruptcy case does not have a statistically, or economically, significant effect on local legal employment (-0.1\%).



        %
        %

        As discussed in Section \ref{sec:Methodology}, the forum shopping channel also plays a crucial role for identification, as it provides an institutional mechanism to corroborate our interpretation of $\gamma_1$. 
        The null effect of forum shopped cases implies that the bankruptcy filing itself is causing the observed impact on legal employment, and not some other confounding factor associated with these bankruptcies. 
        Section \ref{sec:FS_LocalConditions} justifies the validity of using forum shopping as an institutional placebo. 
        
        Finally, the average level of treatment for treated counties is 1.70 bankruptcy filings, 0.47 of which are forum shopped (Table \ref{tab:DescriptiveStats_OnlyBR_FS}).
        Accordingly, the estimates from Column (2) imply that the expected increase in legal employment is roughly 1.2\% among treated county-year observations.

\vskip -0.5 em
        \subsection{Robustness Checks}\label{sec:robust_check}
\vskip -0.5 em
        
        This section reports several robustness exercises aimed at easing concerns regarding our empirical results. We grouped these robustness exercises in three broad categories: (i) sample definition, (ii) controls included, and (iii) key variable measurement.  In all cases our main results are unchanged.
       
        \subsubsection{Sample}\label{sec:robust_check_sample}
\vskip -0.5 em        
        Our results are robust to adjusting the sample considered in the regression analysis along several dimensions. 
        We first 
        include New York and Delaware counties. 
        Second, 
        we drop 1991 from our benchmark sample period.
        Third, 
        we narrow our sample to more relevant subsets of counties: counties where at least one publicly traded firm is located at any time between 1991 and 1996, or counties with at least one publicly traded firm undergoing bankruptcy during our sample period.
        Fourth, 
        we limit the sample to counties where district courts are located.
        Finally, 
        we extend our sample to cover the period 1991-2001 for specification \eqref{eq:spec_bench_BR}. 
       
        \textbf{Include New York and Delaware counties:}
           Our main sample drops counties from New York and Delaware due to the fact that 
           these magnet districts attracted most of the forum shopping cases during the Court Competition Era (which may have impacted their level of legal employment). 
           To evaluate if this restriction is affecting our results, we re-perform our analysis on an augmented sample that includes New York and Delaware counties. Results are reported in Columns (1) and (2) of Table \ref{tab:RBST_IncludeNYDE_Drop1991}. As before, we find that each bankruptcy filing is associated with an increase in local legal employment of almost 1\% and that forum-shopping nullifies this effect. 
           
        \textbf{Restricted Sample Period:}
            Our baseline sample period 1991-1996 can be considered largely free of recessions on an aggregate level. 
            However, it could be argued that the beginning of our sample period coincides with the ending of the 1991 recession (first quarter 1991, \cite{hamilton_2020}).
            To ease concerns about this potential contamination, we drop the year 1991. 
              Columns (3) and (4) of Table \ref{tab:RBST_IncludeNYDE_Drop1991} show that our estimates are largely unaffected by this change. 
        
        \textbf{More Relevant Comparison Groups:}
            One potential issue with our sample is that by including all counties where legal employment is non-zero  
            we also include in our control group counties that could not feasibly experience a bankruptcy shock, as no publicly traded firms were located there.
            %
            %
            We address these concerns by replicating our baseline results with samples that have been restricted to only include counties that could plausibly receive treatment.

            First, 
            we restrict our estimation sample to counties 
            with at least one publicly traded company located therein at any point in the period from 1991 through 1996.
            This restriction significantly reduces our estimation sample to approximately one-third of the observations in our baseline sample. 
            Nonetheless, results in Columns (1) and (2) of Table \ref{tab:RBST_HQ_BR_COURT} confirm all our main findings.
            
            To further make our sample of counties comparable, we restrict our sample to counties that experienced a bankruptcy at some point in our sample period.
            This drastically reduces the sample size to 354 county-year observations.
            Columns (3) and (4) of Table \ref{tab:RBST_HQ_BR_COURT} show our main findings still hold. 
            
        \textbf{Counties with Bankruptcy Courts:}
            Our sample may include a small number of cases where a firm does not forum shop, but nonetheless is still located a substantial distance from their local court. If this is the case, the effect of the bankruptcy may not actually occur in the county the firm is located in, or at the very least could be reduced. To address these potential concerns, we run our benchmark regressions on a sample that only includes 
            counties where a bankruptcy court is located. 
            While there are 90 judicial districts that handle bankruptcies, several of those districts have multiple physical court locations. As a result, there are 
            approximately 
            200 counties that contain a bankruptcy court. 
            Despite greatly reducing the sample size (1074 observations), 
            Columns (5) and (6) of Table \ref{tab:RBST_HQ_BR_COURT} show that our results still hold.
                \footnote{Note that district-year fixed effects are not used because of perfect collinearity for many districts that have only 1 bankruptcy court. State-year then becomes our preferred fixed effect.}
            
        \textbf{Extended Sample Period:}
            To explore whether our results hold when we consider a longer
            period, we rerun our benchmark results from Table \ref{tab:MAIN_90_97_BRonly} for the period 1991-2001. 
            Importantly, these years make up the post-war era's longest business cycle expansion prior to the 2008 financial crisis.  
            The results in Table \ref{tab:RBST_90_02_BRonly} supports our main finding that bankruptcy filings are associated with an increase in local legal employment. 
            
    \subsubsection{Controls}\label{sec:robust_check_controls} 
\vskip -0.5 em 
        In our benchmark specification, we include county level controls for the log of population and the log of the number of employees outside of the legal sector. In this subsection, we extend our set of controls by lagging the controls and by including new time-varying county variables.
       
        \textbf{Lagged Controls:}
        We run our regressions where controls are lagged one year to avoid potential impact of bankruptcies on the control variables.  The results of this estimation are presented in Columns (1) and (2) of Table \ref{tab:RBST_Lag_Unemp_Estab} and they confirm our findings. Note that this approach also avoids potential indirect effects of bankruptcies on the level of employment outside of the legal sector. 
            
        \textbf{Business Cycle:}
        %
        Another possible concern is that our estimates may be sensitive to macroeconomic conditions. 
        Previously we have already addressed this concern in two ways. 
        First, our preferred specification includes district-year fixed effects that fully control for time-varying aggregate shocks at the district and national level.
        Second, in Table \ref{tab:RBST_IncludeNYDE_Drop1991} we drop the year 1991 since it overlaps with the end of a recession. 
        
        Notwithstanding, here we also incorporate additional time-varying controls that could be expected to be correlated with county-level business-cycle 
        fluctuations,
        such as the county-year unemployment rate (from the BLS), and the percent change in the number of establishments outside the legal sector (from the CBP).
        Again, our findings are robust to these alternative specifications.
        These results can be found in Columns (3) through (6) of Table \ref{tab:RBST_Lag_Unemp_Estab}.
        
        

    \subsubsection{Measurement}\label{sec:robust_check_measurement}
\vskip -0.5 em 
        In this section, we explore the robustness of our main results to alternative measurements of bankruptcy, forum shopped bankruptcies, and employment levels in the legal sector and non-legal sector.  In all cases, our findings are aligned with our priors.
        
        \textbf{Binary measurement:}
            Columns (1) and (2) of Table \ref{tab:RBST_Dummy_ValueWeight} show estimates when we replace the bankruptcy and forum shopping count variables with their respective dummies. 
            Our main findings still hold, albeit with reduced significance. 
            This result is expected due to the fact that these measurements for bankruptcy and forum shopping are less precise compared to the benchmark count variables.
            
        \textbf{Value-Weighted Measurement:}
            In our benchmark regressions, we use an issuer-weighted measure of bankruptcy and forum shopping, which simply counts the number of bankruptcy cases.
            This overlooks potential heterogeneity in the impact from bankruptcy shocks of larger versus smaller firms. 
            To explore this, we include regressions where bankruptcy and form shopping are measured by the amount of assets the company had at the time of filing for bankruptcy.
            Columns (3) and (4) of Table \ref{tab:RBST_Dummy_ValueWeight} present these results and confirm our findings for both bankruptcy and forum shopping.
        
        \textbf{Spillover Effects:}
            To address the potential mismeasurement of 
            bankruptcy shocks
            due to spillover effects,  we run an additional specification which controls for potential spillover from nearby bankruptcies. To do this, we create a measure of the number of bankruptcy filings in counties that are \textit{directly adjacent} to county $c$ in year $t$. 
            As shown in Columns (1) and (2) of Table \ref{tab:RBST_BLS_Adj}, we find no evidence that bankruptcies in adjacent counties have a meaningful impact on local legal employment.

        \textbf{Alternative Data Source:}
            To further verify the validity of our results, we replicate our findings by constructing the employment variables from an alternative data source. Specifically, the dependent variable for legal employment, as well as the control for non-legal employment, are constructed using data from the Bureau of Labor Statistics (\BLS) rather than \CBP. 
            The results from Columns (3) and (4) of Table \ref{tab:RBST_BLS_Adj} demonstrate that even with the less complete \BLS \space data we still observe a significant and positive effect from bankruptcies that is canceled out by forum shopping. 

    \subsection{Potential Employment Gains Lost to NY and DE}\label{sec:Pot_Emp_Gain}
\vskip -0.5 em
        So far, our analysis has shown that bankruptcies of publicly listed firms boost the demand of local legal services (Section \ref{sec:legemp}), 
        but only when cases are not forum shopped (Section \ref{sec:legempforsho}).
        Thus, forum shopping is effectively allowing distressed firms to import bankruptcy legal services from 
        one of the two magnet districts: the Delaware and Southern District of New York (NYDE). 
        This section builds of our previous results and uses the estimated effect of bankruptcy and forum shopping 
        from Column (2) in Table \ref{tab:MainEmpLvl} to quantify the unrealized local legal markets gains lost due to firms forum shopping to New York and Delaware.\footnote{
            Table \ref{tab:MainEmpLvl_NYDE_SHOPS} estimates the effect of bankruptcies and forum shopping when only shopping to NYDE is considered. These results are nearly identical to those from Table \ref{tab:MainEmpLvl} verifying that the estimated effects do not change when only forum shopping to NYDE is considered.} 
        
            \begin{figure}[h]
            \begin{subfigure}[b]{0.5\textwidth}
                \centering
                \includegraphics[width = \textwidth]{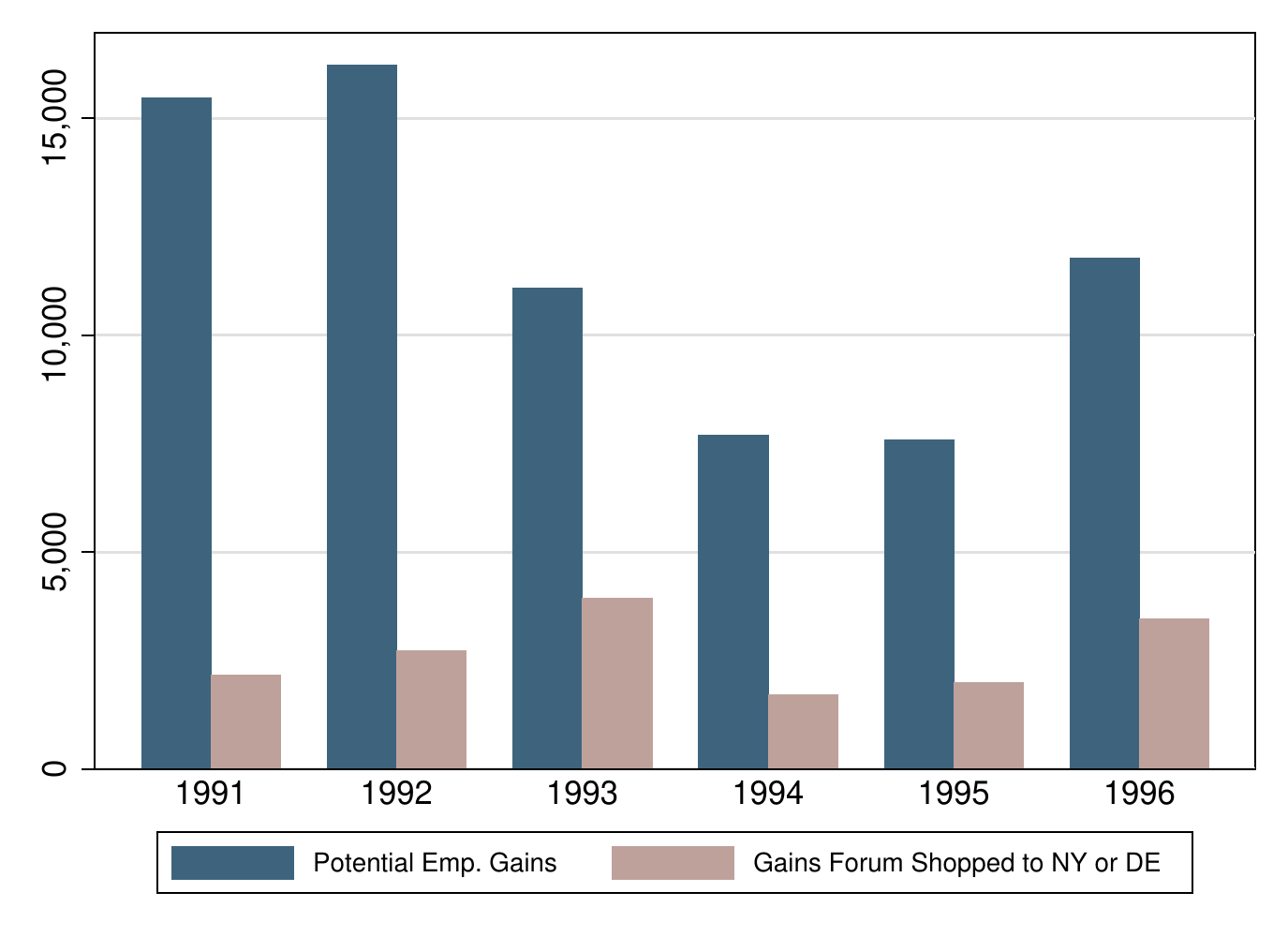}
            \end{subfigure}
            \begin{subfigure}[b]{0.5\textwidth}
                \centering
                \includegraphics[width = \textwidth]{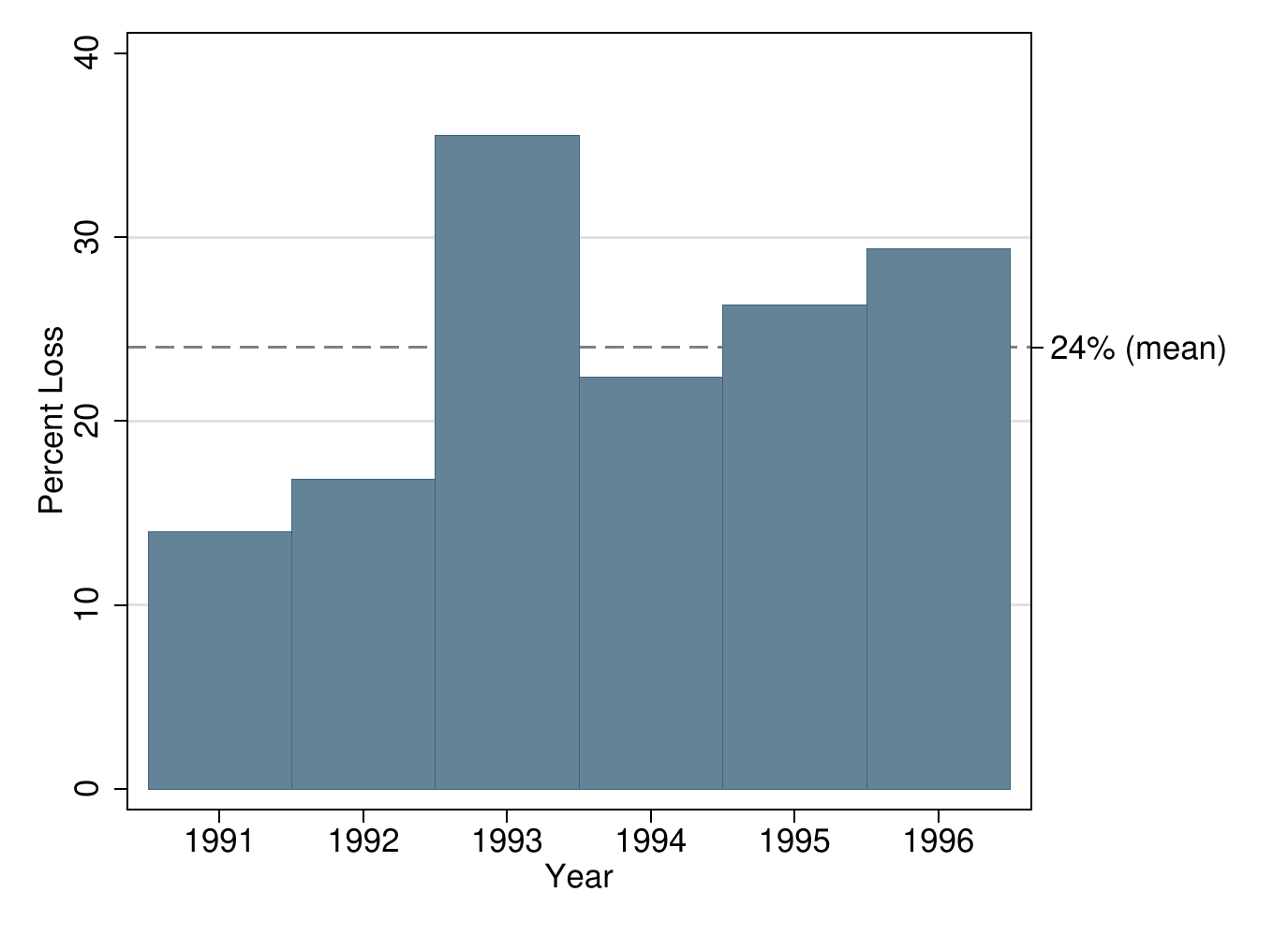}
            \end{subfigure}
            \mycaption{Expected Employment Gains Lost to NY and DE}{\\(left) Comparison of the total predicted potential legal employment gains from Ch11 bankruptcies and the amount of those gains lost to either New York or Delaware via forum shopping.
            \\ (right) The percentage of predicted potential legal employment gains from Ch11 bankruptcy unrealized due to forum shopping to New York or Delaware.}
            \label{FS_Impact_to_gains}
        \end{figure}
        
        We start by estimating the predicted \textit{potential} gains from bankruptcy for each county-year observation if all bankruptcies forum shopped to NYDE had instead been filed locally. 
        To do this, we use the estimated impact of a bankruptcy to calculate the total expected increase in legal employment from bankruptcies that were filed locally \textit{and} the bankruptcies filed in NYDE.
        Next, we use the estimated impact of forum shopping to calculate the amount of these predicted potential gains that are lost to NYDE. 
        Finally, we aggregate these potential gains and losses to show the total amount of potential employment gains being exported to New York and Delaware each year. 
    
        Figure \ref{FS_Impact_to_gains} shows the extent to which potential gains from bankruptcies are being shopped away to New York and Delaware. 
        These losses are considerable.
        Each year, an average of roughly 2,700 jobs, or 24\% of potential gains, are lost from communities affected by major bankruptcies to New York and Delaware.
        Over the court competition era, this adds up to \$1.63 billion in wages in 2021 dollars (\$876 million nominally).
        To put this number in context, 
        \citet{criscuolo_causal_2019} estimate that for government investment in place-based job policies in the U.S. the cost per job created ranges between \$23,000 and \$78,000 in 2021 dollars (\$18,000 and \$63,000 in 2010 dollars). 
        Accordingly, the cost to create those 2,700 lost jobs would be between \$62 and \$211 million in 2021 dollars.
        This back of the envelope calculation indicates that forum shopping may be the source of economically significant costs to local communities, so far unaccounted for by the current policy debate.

\section{A Model of Legal Services Labor Markets}
\vskip -0.8 em


%


This section specializes
a model with skilled and unskilled workers (e.g. \cite{acemoglu_technical_2002}; \cite{Acemoglu_Autor_2011}; \cite{bowlus_wages_2017}) to capture the impact of bankruptcy shocks on local legal labor markets. 
Our theoretical framework 
closely follows \citet{notowidigdo_incidence_2020}.
We consider a dynamic partial equilibrium model of the local legal sector. 
The economy is populated by a mass of heterogeneous households and a representative firm that produces legal services. 
Households differ in their labor productivity: skilled and unskilled. 
Skilled workers can be thought of as lawyers, who require an extremely large investment in training before they can work. 
Whereas unskilled workers can be thought of as para-legals, law clerks, and other legal sector employees who require considerably less upfront costs in training relative to skilled workers.  
The representative firm hires skilled and unskilled workers to produce legal services (the numeraire). 
The demand for legal services is exogenous. 
Labor markets are competitive.

    \subsection{Labor Demand}\label{sec:LaborDemand}
\vskip -0.5 em
    For county $c$ in each period $t$ a profit-maximizing representative firm hires skilled $(\nsct)$ and unskilled $(\nuct)$ workers to produce legal services ($L$) according to the following constant elasticity of substitution production function:
    \begin{equation*}\label{eq:CES}
        L_{c,t} = \theta_{c,t} \Big[ 
        (1-\lambda)(\nuct)^{\zeta} + 
        \lambda(A\nsct)^{\zeta}
        \Big]^{\frac{\alpha}{\zeta}} \;,
    \end{equation*}
    where $\delta \equiv 1/(1-\zeta)$ is the elasticity of substitution between the input factors, $A$ measures the relative efficiency of skilled workers, $\alpha$ is the returns to scale, and $\lambda$ is a share parameter. 
    
    Similar to \citet{notowidigdo_incidence_2020}, $\theta_{c,t}$ is a measure of county level labor demand in the legal service sector. 
    This follows in a similar vein to \citet{bai_demand_2012}, where shocks to demand enter the model as `productivity shocks' through $\theta_{c,t}$.
    In doing so, the model takes into account the fact that even for fixed inputs the level of production will depend on the amount of demand for that good, and in turn the level of demand for the inputs.
    Following our findings in Section \ref{sec:Results}, we use Chapter 11 bankruptcy shocks as a source of plausibly exogenous variation in $\theta_{c,t}$. 

    Markets are competitive and factors receive their marginal product as wages
    \begin{gather*}
        \label{eq:LaborDemandS}
            \wsct = 
            \alpha \theta_{c,t} 
            \Big[ (1-\lambda)(\nuct)^\zeta + \lambda(A\nsct)^\zeta\Big]^{\frac{\alpha - \zeta}{\zeta}}
            \lambda A (A\nsct)^{\zeta - 1}\;, \\
        \label{eq:LaborDemandU}
            \wuct = 
            \alpha \theta_{c,t} 
            \Big[ (1-\lambda)(\nuct)^\zeta + \lambda(A\nsct)^\zeta\Big]^{\frac{\alpha - \zeta}{\zeta}}
            (1-\lambda) (\nuct)^{\zeta - 1}\;.
    \end{gather*}
    To understand the impact of bankruptcy shocks, we take the logarithm and totally differentiate the above expression,
    \begin{gather}
        \label{eq:LaborDemandEvolutionS}
            \Delta \wsct  = 
            \Delta \theta_{c,t} +
            \Big( (\zeta - 1) + (\alpha - \zeta)\pi \Big)\Delta \nsct
            + (\alpha - \zeta)(1 - \pi) \Delta \nuct\;, \\
        \label{eq:LaborDemandEvolutionU}
            \Delta \wuct  = 
            \Delta \theta_{c,t} +
            \Big( (\zeta - 1) + (\alpha - \zeta)(1-\pi) \Big)\Delta \nuct
            + (\alpha - \zeta)(\pi) \Delta \nsct\;,
    \end{gather}
    where $\Delta$ denotes percent change over time and $\pi \equiv \frac{\lambda(A\s)^\zeta}{(1-\lambda)\u^\zeta +\lambda(A\s)^\zeta}$. 
    These expressions describe the dynamics of the equilibrium response of wages to exogenous demand shocks, $\Delta\theta$, in terms of the endogenous response in employment levels, $\Delta\nsct$ and $\Delta\nuct$. 

    \subsection{Labor Supply}\label{sec:LaborSupply}
\vskip -0.5 em
    The households of county $c$ work in the legal sector and differ in their labor productivity type ($j$): skilled  ($j=s$) and unskilled  ($j=u$).
    In each period $t$, households of type $j$ choose consumption $\cjt\in\mathbb{R}_{+}$ and labor hours $\njt\in\mathbb{R}_{+}$ in order to maximize Greenwood–Hercowitz–Huffman preferences
    \begin{equation}\label{eq:Utility}
    \begin{split}
           \underset{\{\cjct,\njct\}^{\infty}_{t=0}}{\max}\quad &
           \sum_{t=0}^{\infty}
           \beta^{t}\frac{1}{1-\sigma}
           \left(\cjct - \frac{(\njct)^{1+\rho_{j}}}{1 + \rho_{j}}\right)^{1-\sigma}
           \qquad \text{s.t.}\; \cjct \leq \wjct \njct\;,
    \end{split}
    \end{equation} 
    where $\rho_{j}>0$ is type $j$'s inverse Frisch elasticity of labor supply, and $\sigma > 0$ decides the agents elasticity of intertemporal substitution ($EIS = \frac{1}{1-\sigma}$). 
    For each hour of work supplied workers of type $j$ receive an hourly wage $\wjt$. 
    Wage compensation is their only source of income.
    
    By totally differentiating the logarithm of the first order conditions, we obtain:
    \begin{multicols}{2}
      \begin{equation}\label{eq:LaborSupplyEvolutionS}
        \Delta \nsct = \frac{1}{\rho_s} \Delta \wsct \;,
      \end{equation}
      \begin{equation}\label{eq:LaborSupplyEvolutionU}
        \Delta \nuct = \frac{1}{\rho_u} \Delta \wuct \;.
      \end{equation}
    \end{multicols}

\noindent\textbf{Equilibrium Definition}
    \textit{An equilibrium is an allocation of consumption and labor across skilled and unskilled households, quantity of legal services, as well as wages for skilled and unskilled workers and price of legal services such that: 
    \textit{(i)} households choose consumption and labor supply to solve \eqref{eq:Utility};
    \textit{(ii)} the representative firm hires workers to maximize its profits;
     \textit{(iii)} labor and legal service output markets clear.}

\section{Structural Estimation and Welfare Analysis}\label{sec:struct_est}
\vskip -0.8 em
This section quantifies the \textit{local} welfare losses experienced by legal workers as a result of large bankruptcy reorganizations being forum shopped away from their local communities.
Following \citet{notowidigdo_incidence_2020}, we estimate the structural parameters of our model using a non-linear simultaneous equations GMM estimator.
Then, we use the estimated model to determine the potential welfare gains associated with moving from the status-quo regime where firms forum shop to a counterfactual regime without forum shopping.






    \subsection{GMM Estimation}\label{sec:GMM_results}
\vskip -0.5 em
    Our model is fully characterized by the wage and employment dynamics described in
    \eqref{eq:LaborDemandEvolutionS}, \eqref{eq:LaborDemandEvolutionU}, \eqref{eq:LaborSupplyEvolutionS}, and \eqref{eq:LaborSupplyEvolutionU}.
    We denote the error terms for these equations as $\Delta\epsilon_{c,t}^{j}$, and rearrange to obtain the following:
    \begin{gather} 
    \scalebox{0.93}[1]{$
        \label{eq:MomentConditionWS}
            \Delta \epsilon_{c,t}^{wS} = 
            \Delta \wsct - 
            \Big[ 
            \Delta \theta_{c,t} +
            \Big( (\zeta - 1) + (\alpha - \zeta)(\pi) \Big)\Delta \nsct
            + (\alpha - \zeta)(1 - \pi) \Delta \nuct
            \Big]\;,$} \\
    \scalebox{0.93}[1]{$
        \label{eq:MomentConditionWU}
            \Delta \epsilon_{c,t}^{wU} = 
            \Delta \wuct - 
            \Big[ 
            \Delta \theta_{c,t} +
            \Big( (\zeta - 1) + (\alpha - \zeta)(1-\pi) \Big)\Delta \nuct
            + (\alpha - \zeta)(\pi) \Delta \nsct
            \Big]\;,$} \\
        \label{eq:MomentConditionNS}
            \Delta \eSct = 
            \Delta \nsct - \frac{1}{\rho_s}\Delta \wsct\;, \\
        \label{eq:MomentConditionNU}
            \Delta \eUct = 
            \Delta \nuct - \frac{1}{\rho_u}\Delta \wuct\;.
    \end{gather}
    The endogenous variables $\Delta\wsct$, $\Delta\wuct$, $\Delta\nsct$, $\Delta\nuct$ depend jointly on each other through this system of simultaneous equations. 
    This creates a potential correlation across the error terms and can result in simultaneity bias when estimating the model parameters. 
    However, bankruptcy shocks create exogenous variation in $\theta_{c,t}$ which implies $\Delta \theta_{c,t}$ is uncorrelated with the error terms. 
    We address the potential simultaneity bias by jointly estimating the model parameters with a GMM procedure using equations  
    \eqref{eq:MomentConditionWS}, \eqref{eq:MomentConditionWU}, \eqref{eq:MomentConditionNS}, \eqref{eq:MomentConditionNU} and instruments derived from the source of exogenous variation in $\Delta\theta$.

  Our model has seven parameters that discipline the labor demand $(\zeta, A, \alpha, \pi, \lambda)$ and labor supply $(\rho_s, \rho_u)$.
    We calibrate four and estimate the rest using \BLS\, data.%
        \footnote{We use \BLS\, data instead of CBP data for two reasons: (\textit{i}) it has additional granularity to help distinguish skilled and unskilled legal workers, (\textit{ii}) it includes data on wages. 
        This comes at the cost of lower geographical coverage. 
        In Section \ref{sec:robust_check_measurement} we show that our results on the impact of bankruptcies on legal employment are similar to the ones we obtain using CBP data.} 
    The production parameters $A$, $\zeta$, $\pi$, and $\lambda$ are calibrated as in \citet{notowidigdo_incidence_2020}.
    Let 
     $\lambda = \frac{
    (1-\mu)^{\zeta-1}}
    {(A_s \mu)^{\zeta-1} + (1-\mu)^{\zeta-1}}$ and let $\mu=0.42$ denote the average share of skilled workers in the legal labor market.%
        \footnote{We estimate skilled employment using NAICS 54111 `Offices of Lawyers' and dividing by 2.3 using the estimate that there are 1.3 non-lawyers for every lawyer as suggested by \citet{freedman_how_2005}. We estimate unskilled employment using NAICS 54119 plus the portion of unskilled removed from NAICS 54111. Data is from 1990 through 2001, averaged across time. \textit{Source:} \href{https://www.bls.gov/cew/downloadable-data-files.htm}{BLS}.\nocite{wise_qcew_nodate}
        }
        Then, $A=2.40$ equals the wage premium, computed as the ratio of the
    average wages of skilled and unskilled workers.\footnote{ 
        We use the earliest available detailed occupational wage data, which is from 2000. Average wage for unskilled uses data for paralegals, legal assistants, court reporters, and law clerks. Skilled uses data for lawyers. \textit{Source:} \href{https://www.bls.gov/oes/bulletin_2000.pdf}{BLS}.
        \nocite{BLS_occupational_wage_2000}
    }
    Following a large body of literature, we set the elasticity of substitution equal to the estimate from \citet{katz_changes_1992} of $1.4$, which implies $\zeta = 0.29$. Accordingly, $\pi = 0.63$.
    We estimate the returns to scale $\alpha$, and the Frisch elasticities of substitution $1/\rho_s$ and $1/\rho_u$ using GMM.

    We use Chapter 11 bankruptcy shocks to construct the exogenous demand shocks $\Delta \theta_{c,t}$. Let the demand for legal services for county $c$ in period $t$ be 
    \[ \theta_{c,t} = \bar{L}_{c} + \Phi \cdot \BR_{c,t} \;. \]
    where $\BR_{c,t}$ denotes bankruptcy cases filed locally, $\bar{L}_c$ denotes the county's `steady state' level of demand for legal services and $\Phi$ is the per-bankruptcy expected increase in legal expenses. 
    By taking the logarithm and differentiating, we obtain 
    \begin{equation*}
        \Delta\theta_{c,t} = \frac{\Phi \cdot (\BR_{c,t} - \BR_{c,t-1})} {\bar{L}_c + \BR_{c,t}} \;.
    \end{equation*}
    We calculate $\Phi$ as the average annual legal fees for a Chapter 11 bankruptcy using \LP. We compute $\bar{L}_{c}$ by multiplying county level GDP data for all professional and legal services (NAICS 54)\footnote{
        We collect county level real GDP for NAICS 54, Professional and Business services during 2003. We use 2003 because it is a representative year not impacted by bankruptcy shocks in our sample and data was not available prior to 2000. \textit{Source:} \href{https://apps.bea.gov/itable/iTable.cfm?ReqID=70&step=1&acrdn=5}{BEA}. \nocite{BEA_county_gdp_2003}} for the average proportion of employment in that sector attributed to only the legal industry (NAICS 5411).
    
    In our estimation, we use instrumental variables based on the two sources of exogenous variation in $\Delta\theta_{c,t}$: the contemporaneous non-forum shopped bankruptcies ($\BR_{c,t}$) and the non-forum shopped bankruptcy in the previous period ($\BR_{c,t-1}$). 
    To achieve identification, we also include non-linear functions of these variables. 
    Our preferred specification (Table \ref{tab:gmm_results}) uses combinations of $\{BR_{c,t},(\BR_{c,t-1}),(\BR_{c,t-1})^2, $
    $(\BR_{c,t-1})^3, \BR_{c,t-1})^4\}$ and a constant as instrumental variables resulting in 14 moment conditions. 
    We select our preferred specification using the usual $\chi^2$ over-identification test.
    Table \ref{tab:gmm_results_appendix} in Appendix \ref{App:GMM} shows our results are supported by a variety of other choices of instruments as well.  

    We use \BLS\; data to measure employment levels and wages for skilled and unskilled labor in the legal sector. 
    As before, we omit New York and Delaware from our estimation sample. However, due to limited data availability, we expand our sample through the year 2001. Limiting the sample to only the Court Competition Era (1991-1996) yields very similar estimates albeit with lower significance due to the limited number of observations (Table \ref{tab:gmm_results_appendix}). 
    More information on data and sensitivity to sample selection can be found in Appendix \ref{App:GMM}.


    \begin{table}[H]
    \setlength{\tabcolsep}{17pt}
        \caption{GMM Estimates of model parameters.}
        \label{tab:gmm_results}
        \begin{adjustwidth}{-2in}{-2in}
            \centering
            \begin{threeparttable}
                {
                \begin{tabular}{l*{4}{c}}
                    \toprule
                    &\multicolumn{1}{c}{$1/\rho_s$}    &\multicolumn{1}{c}{$1/\rho_u$}    &\multicolumn{1}{c}{$\alpha$}     &\multicolumn{1}{c}{Observations}\\

                    \midrule
                    Estimates  &      0.0853\sym{***}&       0.264\sym{**} &       1.293\sym{***} & 2370 \\
                                &    (0.0307)         &     (0.120)         &     (0.150)          &  \\

                    \bottomrule
                    \multicolumn{4}{l}{\footnotesize Standard errors clustered at county level in parentheses}\\
                    \multicolumn{4}{l}{\footnotesize \sym{*} \(p<0.1\), \sym{**} \(p<0.05\), \sym{***} \(p<0.01\)}\\    
                \end{tabular}
                }
            \begin{tablenotes}
                \item [] \footnotesize \textit{Note:}
                Instruments for $\Delta \epsilon_{c,t}^{wS}$ and $\Delta \epsilon_{c,t}^{wU}$: \{$(\BR_{c,t-1})$, $(\BR_{c,t-1})^2$, $(BR_{c,t-1})^3$, $(\BR_{c,t-1})^4$\}.
                Instruments for $\Delta \eSct$ and $\Delta \eUct$: \{$\BR_{c,t}$, $(\BR_{c,t-1})$\}, constant.
                Wages and unemployment are detrended prior to estimation using county-year population, log of county-year employment in all non-legal sectors with fixed effects for \countyDisY. \textit{Sample:} \sampleNYDE. \textit{Period:} \sampleyearsExtendedPeriod. \textit{Sources:} \sourceAVGWAGE. 
            \end{tablenotes}
            \end{threeparttable}
        \end{adjustwidth}
    \end{table}

     Table \ref{tab:gmm_results} presents the results of our GMM estimation.
    As expected, we estimate the Frisch elasticity for the extensive margin to be lower for skilled labor ($1/\rho_s$) than it is for unskilled labor ($1/\rho_u$). 
    These estimates are well within the 0 to 1 range found by much of the micro literature (\citealp{whalen_estimates_2017}), and are quite similar to estimates for the extensive margin that range from 0.01 to 0.7 (e.g. \citealp{bianchi_icelands_2001}; \citealp{brown_link_2013}).\footnote{
        For the Frisch Elasticity of the extensive margin \citet{bianchi_icelands_2001} estimates 0.4 for the whole labor force, ranging from 0.1 for women and 0.6 for men; \citet{brown_link_2013} estimates range from 0.01 to 0.17 using workers near retirement age.}
    The estimated returns to scale parameter $\alpha$ is greater than one implying increasing returns to scale. However, the two-sided 95\% confidence interval has a lower bound of 0.999 so we fail to reject the possibility of constant returns to scale.
    The over-identification test for this model does not reject the null hypothesis that the empirical moment conditions' deviations from 0 are due to chance ($p$ = 0.321).
    Finally, Figure \ref{fig:GMM_fit} in Appendix \ref{App:Figures} demonstrates that our calibrated model predicts reasonably well the observed outcomes.

    \subsection{Welfare Analysis}
\vskip -0.8 em
    With our fully calibrated model, we perform counterfactual welfare analysis to estimate the losses to workers caused by forum shopping. 
    To do so, we measure the increase in consumption that would have made skilled and unskilled households indifferent to a regime where all forum shopped bankruptcies were instead filed locally.
    For a household of a given labor type, the consumption equivalent variation implied by our model is defined as $\epsilon$ from the expression
    \begin{equation}
    \scalebox{0.95}[1]{$
        \sum_{t=0}^{T}
        \beta^{t}\frac{1}{1-\sigma}
        \left( 
        \FSct(1 + \epsilon) - \frac{(\FSnt)^{1+\rho_{j}}}{1 + \rho_{j}}
        \right)^{1-\sigma}
        =       
        \sum_{t=0}^{T}
        \beta^{t}\frac{1}{1-\sigma}
        \left( 
        \NFSct - \frac{(\NFSnt)^{1+\rho_{j}}}{1 + \rho_{j}}
        \right)^{1-\sigma}\;,$}
    \end{equation}
    where $\NFSct$ and $\NFSnt$ denote the predicted levels of consumption and labor under the counterfactual regime with no forum shopping.

    \begin{table}[H]
        \setlength{\tabcolsep}{10pt}
        \caption{Consumption Equivalent Variation}
        \label{tab:EquivalentVariation}
        \begin{adjustwidth}{-2in}{-2in}
            \centering
            \begin{threeparttable}
                {
                \begin{tabular}{l*{5}{c}}
                    \toprule
                    &\multicolumn{2}{c}{($\sigma = 2$, $EIS=0.5$)}
                    &&\multicolumn{2}{c}{($\sigma = 0.5$, $EIS=2$)}\\
                    
                    
                    &\multicolumn{1}{c}{Skilled}    
                    &\multicolumn{1}{c}{Unskilled}    
                    &&\multicolumn{1}{c}{Skilled}     
                    &\multicolumn{1}{c}{Unskilled}\\
                    \cmidrule{2-3}\cmidrule{5-6}
                    & 0.74\%
                    & 0.77\%
                    && 0.79\%
                    & 0.82\% \\
                    \bottomrule
                \end{tabular}
                }
            \end{threeparttable}
        \end{adjustwidth}
    \end{table}
    
    Table \ref{tab:EquivalentVariation} presents our findings for the consumption equivalent variation of both worker types averaged across all counties that experienced non-forum-shopped bankruptcy during the sample period. 
    In line with the literature we present our results for a $\sigma$ of 0.5 and 2 (e.g. \cite{havranek_measuring_2015}; \citet{karp_selfish_2021}). 
    %
    In either case, we find that both types of households would need a nearly one percent increase in their consumption level \textit{per year} in order to achieve the same level of utility they otherwise would have without forum shopping. 
    We also find that forum shopping has a larger effect on unskilled workers, who make up the majority of the labor force. 
    
    To the best of our knowledge, this is the first attempt to construct a welfare measure of the impact of forum shopping. 
    These estimates from our parsimonious model show that moving to a regime without forum shopping can have economically significant welfare gains for workers in the legal service industry.  
\section{Conclusion}\label{conclusion}
\vskip -0.8 em




Fluctuations in the financial health of major employers play a critical role in the well-being of local labor markets.
Much to the detriment of these local economies, when a large corporation files for bankruptcy employment falls (\cite{bernstein_bankruptcy_2019}). 
However, the adverse effects of bankruptcy shocks on overall local employment hides the differential impact across sectors in the local production network (\cite{carvalho_micro_2014}).
The legal sector in the bankrupt firms' local area may indeed benefit from the increased demand for legal services, 
a substantial transfer overlooked by the previous literature.

In this paper, we study empirically and theoretically the effect of Chapter 11 bankruptcy shocks on local legal labor markets. 
After constructing a novel database of Chapter 11 bankruptcy reorganizations, we find that each Chapter 11 bankruptcy reorganization of a publicly traded firm is associated with a 1\% increase in county legal employment for each year of a bankruptcy. 
Back-of-the-envelope calculations estimate that forum shopping \textit{exported away} 24\% of the total potential employment gains from local communities with distressed firms. 

We develop a parsimonious model of the legal service sector to structurally estimate the local welfare losses of legal sector workers in counties that experience bankruptcy shocks. Using our calibrated model, we estimate the consumption equivalent variation implied by moving from a world with forum shopping to a world without it. We find the local welfare losses to be roughly 1\% for workers in the legal sector. 

We focus our analysis on the Court Competition Era, a period from 1991 to 1996 that witnessed a boom in forum shopping.
From a methodological point of view, we propose a novel identification approach that uses forum shopping as institutional placebo to create variation in the location of the filing court that are exogenous to local economic conditions.
It is worth mentioning that forum shopping is not exclusive to bankruptcy reorganization procedures. For instance, patent laws allow for the possibility of forum shopping patent litigation cases. 
While the institutional details may be different, our identification approach may prove useful in that context, or other similar contexts, as well.

Our analysis uncovers economically significant costs to local communities from forum shopping. 
These costs have not been accounted for in either the recent policy debate around the Bankruptcy Venue Reform Act bill of 2021, or the ongoing debate in the legal literature over the efficiency of forum shopping.
While these estimates may only be a preliminary look into the potentially large costs of forum shopping today, our analysis suggests that a deeper look at the production network of large corporate bankruptcies is necessary to fully understand the costs to local communities in a modern setting.

\setlength\bibitemsep{0.1pt}
{
\singlespacing
\printbibliography
}

\pagebreak
\addappheadtotoc 
\appendix
\appendixpage

\section{Summary Stats}\label{App:SumStat}
    {\colsepx
    \begin{table}[H]
        \centering
        \begin{threeparttable}[H]
            \caption{Descriptive Statistics for County Level Economic Measures.}
            {
\def\sym#1{\ifmmode^{#1}\else\(^{#1}\)\fi}
\begin{tabular}{l*{2}{cc}}
\toprule
                    &\multicolumn{2}{c}{\textbf{Full Sample}}&\multicolumn{2}{c}{\textbf{County-Year with BR}}\\
                    &        Mean&   Std. Dev.&        Mean&   Std. Dev.\\
\midrule
Legal Emp           &         305&        1661&        5029&        7761\\
\addlinespace
County Pop.         &       88769&      281261&      915621&     1240253\\
\addlinespace
Employment Non-Legal&       32785&      118582&      406210&      510300\\
\midrule
Observations        &       16058&            &         451&            \\
\bottomrule
\end{tabular}
}

            \label{tab:Extra_Desc_Stats}
            \begin{tablenotes}
                \footnotesize\textit{Note:} County data descriptive statistics for employment in the legal sector, county population, employment in all sectors besides legal services. \textit{Sources:} \source.\\
                    \item \textbf{Full Sample}: All counties with non-zero legal employment during the period 1991-1996, not including counties from New York and Delaware.\\         
                    \item \textbf{County-Year with BR}: All counties where at least one bankruptcy took place during the period 1991-1996, not including counties from New York and Delaware. 
            \end{tablenotes}
        \end{threeparttable}
    \end{table}
    }

\section{Figures}\label{App:Figures}
\setcounter{figure}{2}
\begin{figure}[H]
    \centering
    \includegraphics[width =\textwidth]{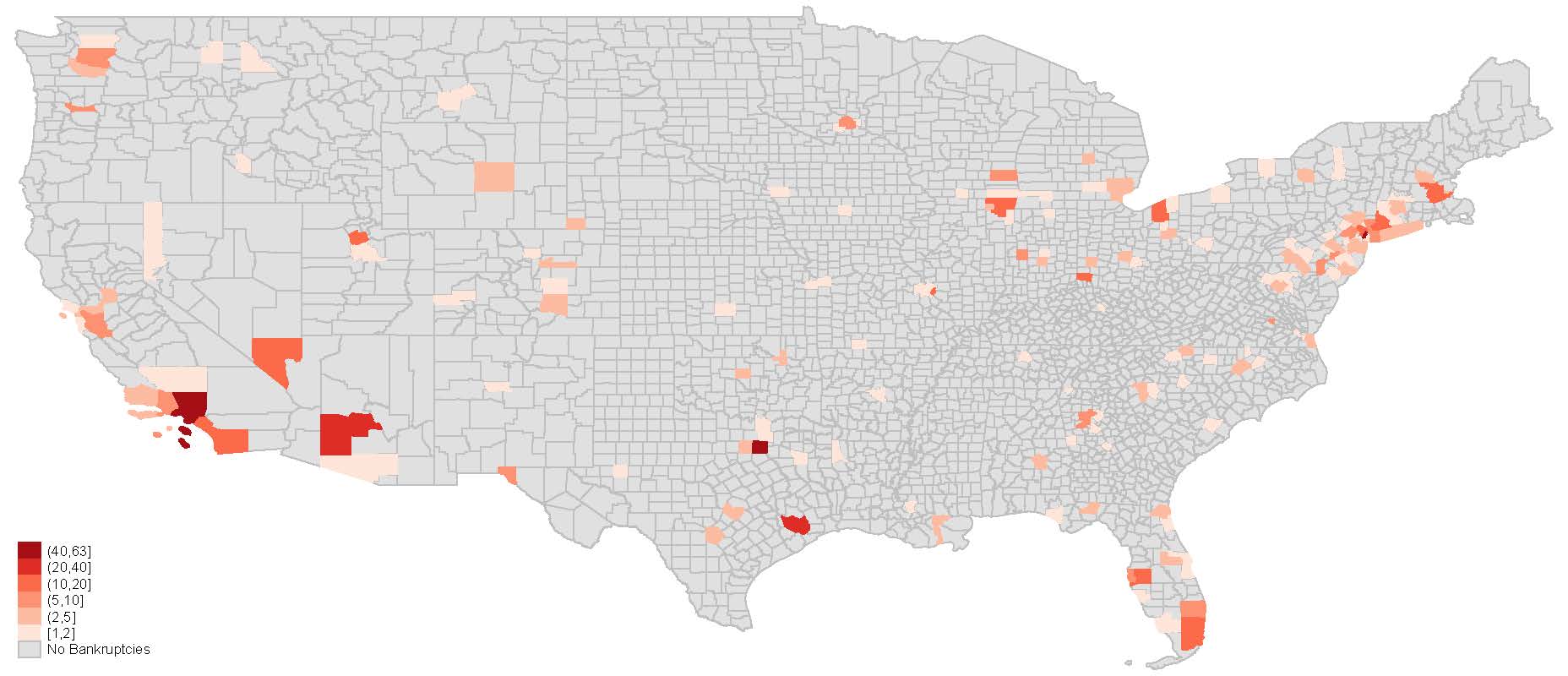}
    \caption{\footnotesize Number of Ch11 Bankruptcies in each county from 1991 to 1996.}
    \label{BR_County_Map}
\end{figure}


\begin{figure}[H]
    \centering
    \includegraphics[width =\textwidth]{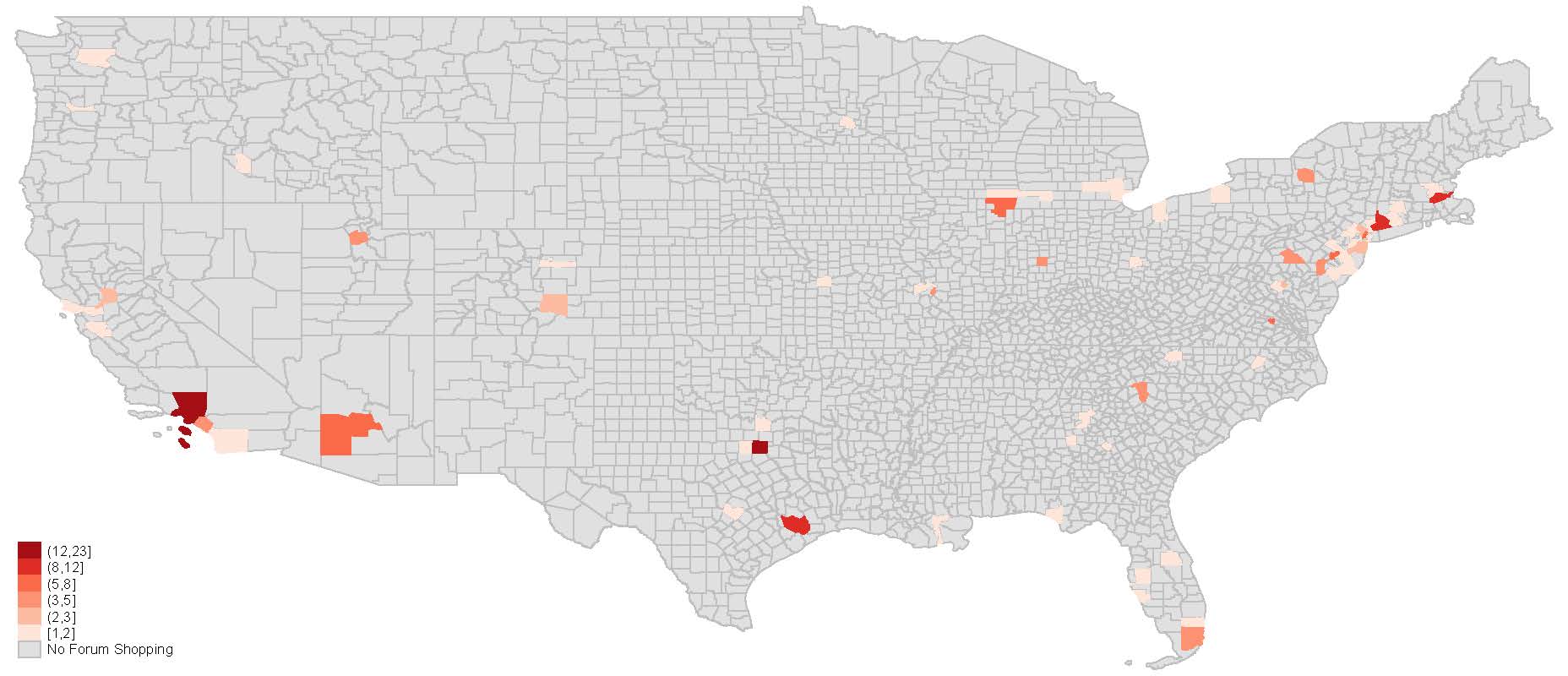}
    \caption{\footnotesize Number of Forum Shopped Ch11 Bankruptcies in each county from 1991 to 1996.}
    \label{FS_County_Map}
\end{figure}

\setcounter{figure}{6}
\begin{figure}[H]
    \begin{subfigure}[b]{0.5\textwidth}
        \centering
        \includegraphics[width = \textwidth, height = 4.3cm]{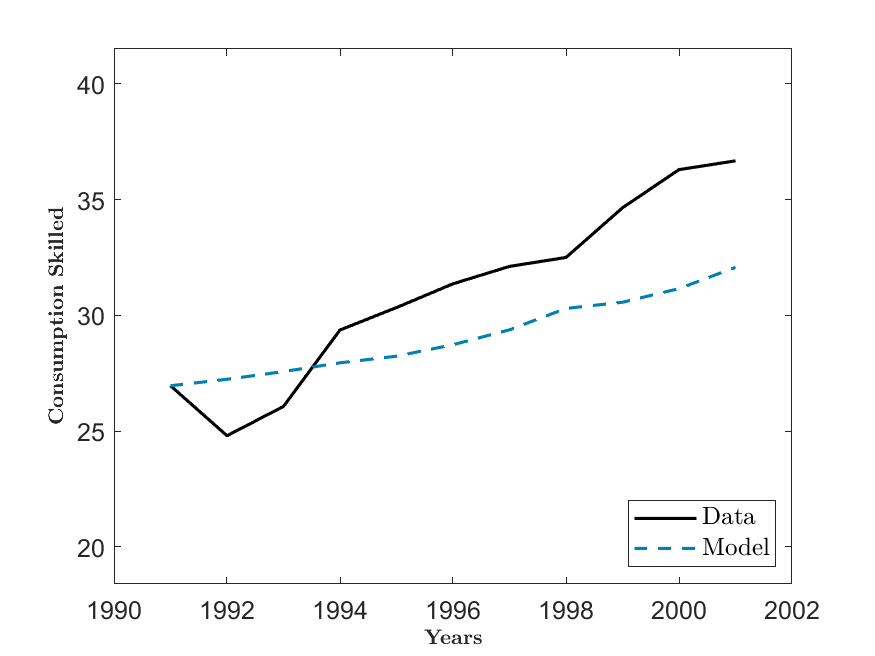}
    \end{subfigure}
    \begin{subfigure}[b]{0.5\textwidth}
        \centering
        \includegraphics[width = \textwidth, height = 4.3cm]{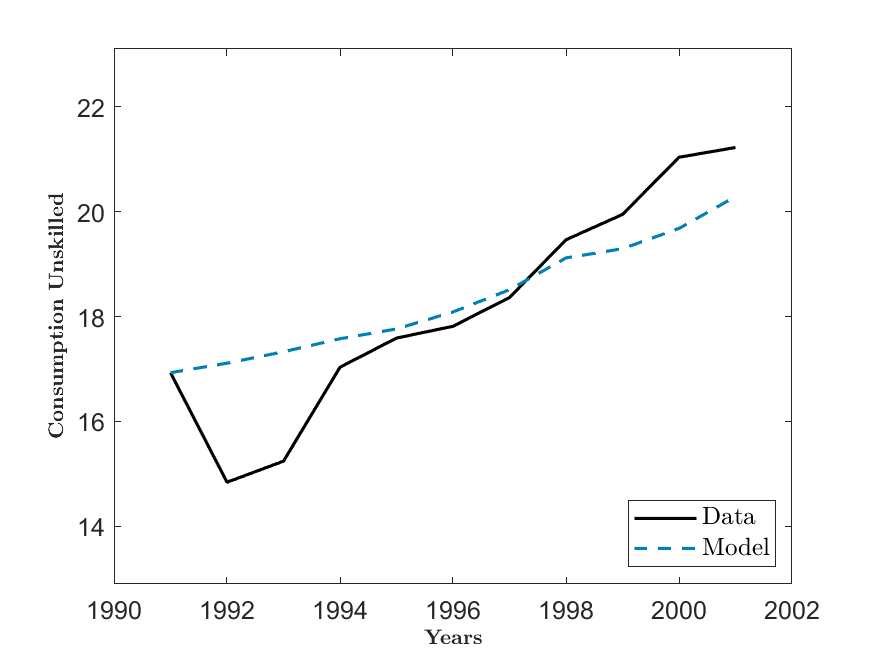}
    \end{subfigure}
    \mycaption{Calibrated Model Predicted Values vs. Observed Values}{Comparison of predicted consumption levels to the observed levels in the data for counties who experience at least one non-forum shopped bankruptcy shock during the estimation period.}
    \label{fig:GMM_fit}
\end{figure}

\section{Placebo Tests}\label{App:Placebo}
    Here we present an additional placebo exercise to accompany Section \ref{sec:Placebo}. Table \ref{tab:placebo_DUMMY} presents placebo regression using an binary indicator for bankruptcy as opposed to the count measure used in Table \ref{tab:placebo}.
    {\colsepx
    \begin{table}[H]
            \adjustbox{max width=\textwidth, center = \textwidth}{%
            \begin{threeparttable}
            \caption{Employment Level Regressions with lags for dummy indicator of bankruptcy shock.}
            
            {
\def\sym#1{\ifmmode^{#1}\else\(^{#1}\)\fi}
\begin{tabular}{l*{4}{c}}
\toprule
                    &\multicolumn{1}{c}{(1)}         &\multicolumn{1}{c}{(2)}         &\multicolumn{1}{c}{(3)}         &\multicolumn{1}{c}{(4)}         \\
\midrule
L1.Bankruptcy       &     0.00368         &                     &                     &     0.00413         \\
                    &   (0.00694)         &                     &                     &   (0.00679)         \\
\addlinespace
L2.Bankruptcy       &                     &    -0.00160         &                     &    -0.00205         \\
                    &                     &   (0.00606)         &                     &   (0.00581)         \\
\addlinespace
L3.Bankruptcy       &                     &                     &     0.00109         &     0.00181         \\
                    &                     &                     &   (0.00810)         &   (0.00815)         \\
\addlinespace
ln(County Population)&       0.715\sym{***}&       0.716\sym{***}&       0.715\sym{***}&       0.715\sym{***}\\
                    &     (0.144)         &     (0.144)         &     (0.144)         &     (0.144)         \\
\addlinespace
ln(Emp Non-Legal)   &       0.170\sym{***}&       0.170\sym{***}&       0.171\sym{***}&       0.171\sym{***}\\
                    &    (0.0489)         &    (0.0489)         &    (0.0490)         &    (0.0490)         \\
\midrule
County FE           &         Yes         &         Yes         &         Yes         &         Yes         \\
District-Year FE    &         Yes         &         Yes         &         Yes         &         Yes         \\
R-Squared           &       0.984         &       0.984         &       0.984         &       0.984         \\
Observations        &       15931         &       15931         &       15930         &       15930         \\
\bottomrule
\multicolumn{5}{l}{\footnotesize Standard errors clustered at county level in parentheses}\\
\multicolumn{5}{l}{\footnotesize \sym{*} \(p<0.1\), \sym{**} \(p<0.05\), \sym{***} \(p<0.01\)}\\
\end{tabular}
}

            \label{tab:placebo_DUMMY}
            
            \begin{tablenotes}
                \item [] \footnotesize\textit{Note:} \tabnote{\lnLegEmp}{\regressorsPLACEBOdummy}{\countyDisY}{\controls}{\clustercounty}{\samplerestr}{\sampleyears}{\source}
            \end{tablenotes}
            \end{threeparttable}
            }
        \end{table}
    }


    
    

    \section{Fixed Effects}\label{App:Fixed_Effects}
    This section reports the results for our main forum shopping regressions across a variety of different fixed effect specifications. 
    Table \ref{tab:FE_County_Year_RegY_DivY_SY_DisY} reports our forum shopping regressions- as in specification \eqref{eq:spec_bench_BR_FS}- for each of the fixed effects used in Table \ref{tab:MAIN_90_97_BRonly}: county, county and year, county and region-year, county and division-year, county and state-year, county and district-year.
    In each case, they confirm our results and show that the combined effect of a forum shopped bankruptcy is not statistically different from 0.
    %
    %
    Recall that Column (2) of Table \ref{tab:MainEmpLvl} reports these results for county and district-year fixed effects, which is our preferred specification and the most stringent fixed effects. 
    Note that for completeness the estimates for the controls are also included here, although they are omitted in the main body of the paper.

    \begin{table}[h]
            \adjustbox{max width=\textwidth, center = \textwidth}{%
            \begin{threeparttable}
            \caption{Employment Level Regressions}
            
            {
\def\sym#1{\ifmmode^{#1}\else\(^{#1}\)\fi}
\begin{tabular}{l*{6}{c}}
\toprule
                    &\multicolumn{1}{c}{(1)}         &\multicolumn{1}{c}{(2)}         &\multicolumn{1}{c}{(3)}         &\multicolumn{1}{c}{(4)}         &\multicolumn{1}{c}{(5)}         &\multicolumn{1}{c}{(6)}         \\
\midrule
Bankruptcies        &     0.00948\sym{***}&      0.0101\sym{***}&      0.0100\sym{***}&     0.00862\sym{**} &     0.00878\sym{**} &      0.0100\sym{***}\\
                    &   (0.00349)         &   (0.00355)         &   (0.00356)         &   (0.00348)         &   (0.00385)         &   (0.00373)         \\
\addlinespace
Forum Shopping      &    -0.00741         &    -0.00873\sym{*}  &    -0.00689         &    -0.00772         &    -0.00759         &     -0.0111\sym{*}  \\
                    &   (0.00518)         &   (0.00518)         &   (0.00522)         &   (0.00527)         &   (0.00623)         &   (0.00637)         \\
\addlinespace
ln(County Population)&       0.797\sym{***}&       0.750\sym{***}&       0.694\sym{***}&       0.651\sym{***}&       0.726\sym{***}&       0.715\sym{***}\\
                    &     (0.115)         &     (0.125)         &     (0.134)         &     (0.134)         &     (0.137)         &     (0.144)         \\
\addlinespace
ln(Emp Non-Legal)   &       0.182\sym{***}&       0.166\sym{***}&       0.170\sym{***}&       0.174\sym{***}&       0.167\sym{***}&       0.170\sym{***}\\
                    &    (0.0418)         &    (0.0468)         &    (0.0474)         &    (0.0479)         &    (0.0481)         &    (0.0489)         \\
\midrule
                    &                     &                     &                     &                     &                     &                     \\
Fixed Effects       &                     &                     &                     &                     &                     &                     \\
County              &         Yes         &         Yes         &         Yes         &         Yes         &         Yes         &         Yes         \\
Year                &          No         &         Yes         &          No         &          No         &          No         &          No         \\
Region-Year         &          No         &          No         &         Yes         &          No         &          No         &          No         \\
Division-Year       &          No         &          No         &          No         &         Yes         &          No         &          No         \\
State-Year          &          No         &          No         &          No         &          No         &         Yes         &          No         \\
District-Year       &          No         &          No         &          No         &          No         &          No         &         Yes         \\
\midrule            &                     &                     &                     &                     &                     &                     \\
R-Squared           &       0.984         &       0.984         &       0.984         &       0.984         &       0.984         &       0.984         \\
Observations        &       15963         &       15963         &       15963         &       15963         &       15963         &       15931         \\
\bottomrule
\multicolumn{7}{l}{\footnotesize Standard errors clustered at county level in parentheses}\\
\multicolumn{7}{l}{\footnotesize \sym{*} \(p<0.1\), \sym{**} \(p<0.05\), \sym{***} \(p<0.01\)}\\
\end{tabular}
}

            \label{tab:FE_County_Year_RegY_DivY_SY_DisY}
            
            \begin{tablenotes}
                \item [] \footnotesize\textit{Note:} \tabnote{\lnLegEmp}{\regressorsNoShock}{\FEall}{\controls}{\clustercounty}{\samplerestr}{\sampleyearsExtendedPeriod \sampleNYDE}{\source}
            \end{tablenotes}
            \end{threeparttable}
            }
        \end{table} 

\section{NYDE Forum Shopping}\label{App:NYDE_FS}
    In Section \ref{sec:Pot_Emp_Gain} we are interested in estimating the number of jobs being lost due to forum shopping to New York and Delaware from local communities affected by bankruptcies. 
    In this section, we verify that the estimated effect of forum shopping does not change when only forum shopping to New York and Delaware is considered. 
    Table \ref{tab:MainEmpLvl_NYDE_SHOPS} confirms this conjecture, as the estimated effect of bankruptcy and forum shopping is nearly identical to Table \ref{tab:MainEmpLvl}.

    {
    \colsepx
    \begin{table}[h]
            \adjustbox{max width=\textwidth, center = \textwidth}{%
            \begin{threeparttable}
            \caption{Employment Level Regressions with Forum Shopping Only to NY or DE}
            
            {
\def\sym#1{\ifmmode^{#1}\else\(^{#1}\)\fi}
\begin{tabular}{l*{3}{c}}
\toprule
                    &\multicolumn{1}{c}{(1)}         &\multicolumn{1}{c}{(2)}         &\multicolumn{1}{c}{(3)}         \\
\midrule
Bankruptcies        &     0.00791\sym{**} &      0.0100\sym{***}&                     \\
                    &   (0.00334)         &   (0.00373)         &                     \\
\addlinespace
Non-FS Bankruptcies &                     &                     &      0.0100\sym{***}\\
                    &                     &                     &   (0.00373)         \\
\addlinespace
Forum Shopping to NYDE&                     &     -0.0126\sym{*}  &    -0.00252         \\
                    &                     &   (0.00659)         &   (0.00560)         \\
\midrule
County FE           &         Yes         &         Yes         &         Yes         \\
District-Year FE    &         Yes         &         Yes         &         Yes         \\
R-Squared           &       0.984         &       0.984         &       0.984         \\
Observations        &       15931         &       15931         &       15931         \\
\bottomrule
\multicolumn{4}{l}{\footnotesize Standard errors clustered at county level in parentheses}\\
\multicolumn{4}{l}{\footnotesize \sym{*} \(p<0.1\), \sym{**} \(p<0.05\), \sym{***} \(p<0.01\)}\\
\end{tabular}
}

            \label{tab:MainEmpLvl_NYDE_SHOPS}
            
            \begin{tablenotes}
                \item [] \footnotesize\textit{Note:} \tabnote{\lnLegEmp}{\regressorsNYDESHOP}{\countyDisY}{\controls}{\clustercounty}{\samplerestr}{\sampleyears}{\source}
            \end{tablenotes}
            \end{threeparttable}
            }
        \end{table}
    }

\section{Robustness Checks}\label{App:Robust_Check}
    This section reports several robustness exercises discussed in Section \ref{sec:robust_check} aimed at addressing potential concerns regarding our empirical results. We group these robustness exercises in three broad categories: (i) sample definition, (ii) controls included, and (iii) key variable measurement.  In all cases our main results are unchanged.

    \subsection{Sample}

        In this subsection, we show that our results are robust to adjusting our sample considered in the regression analysis along several dimensions.
        First, Table \ref{tab:RBST_IncludeNYDE_Drop1991} adjusts the sample by including New York and Delaware counties data in Columns (1) and (2) then by dropping 1991 from our benchmark sample period in Columns (3) and (4).
        Second, Table \ref{tab:RBST_HQ_BR_COURT} limits the sample to more relevant counties by only including: \textit{(i)} counties where a publicly traded firm was located at any time between 1991 and 1996, \textit{(ii)} counties where at least one bankruptcy took place during our sample period (1991-1996), \textit{(iii)} counties where district courts are located. 
        Note that for the disctrict court sample (iii) we use state-year fixed effects due to perfect colinearity for many districts that have only 1 bankruptcy court.
        Finally in Table \ref{tab:RBST_90_02_BRonly} we extend our sample to cover the period 1991-2001, we do not include forum shoppping because this is outside of the Court Competition Era.
        See Section \ref{sec:robust_check_sample} for a detailed discussion.
        
        {\colsepx
        \begin{table}[H]
            \adjustbox{max width=\textwidth, center = \textwidth}{%
            \begin{threeparttable}
            \caption{Employment Level Regressions}
            
            {
\def\sym#1{\ifmmode^{#1}\else\(^{#1}\)\fi}
\begin{tabular}{l*{4}{c}}
\toprule
                    &\multicolumn{2}{c}{\textbf{Include NYDE}}  &\multicolumn{2}{c}{\textbf{Drop 1991}}     \\
                    &\multicolumn{1}{c}{(1)}         &\multicolumn{1}{c}{(2)}         &\multicolumn{1}{c}{(3)}         &\multicolumn{1}{c}{(4)}         \\
\midrule
Bankruptcies        &     0.00669\sym{**} &     0.00836\sym{**} &     0.00831\sym{**} &      0.0111\sym{**} \\
                    &   (0.00308)         &   (0.00335)         &   (0.00410)         &   (0.00478)         \\
\addlinespace
Forum Shopping      &                     &    -0.00856         &                     &     -0.0118\sym{*}  \\
                    &                     &   (0.00577)         &                     &   (0.00702)         \\
\midrule
County FE           &         Yes         &         Yes         &         Yes         &         Yes         \\
District-Year FE    &         Yes         &         Yes         &         Yes         &         Yes         \\
R-Squared           &       0.985         &       0.985         &       0.986         &       0.986         \\
Observations        &       16314         &       16314         &       13327         &       13327         \\
\bottomrule
\multicolumn{5}{l}{\footnotesize Standard errors clustered at county level in parentheses}\\
\multicolumn{5}{l}{\footnotesize \sym{*} \(p<0.1\), \sym{**} \(p<0.05\), \sym{***} \(p<0.01\)}\\
\end{tabular}
}

            \label{tab:RBST_IncludeNYDE_Drop1991}
            
            \begin{tablenotes}
                \item [] \footnotesize\textit{Note:} \tabnote{\lnLegEmp}{\regressorsNoShock}{\countyDisY}{\controls}{\clustercounty}{Columns (1) and (2) \sampleNYDE; Columns (3) and (4) \samplerestrRBST}{Columns (1) and (2) \sampleyears; Columns (3) and (4) \sampleyearsDROPNintyOne}{\source}
            \end{tablenotes}
            \end{threeparttable}
            }
        \end{table} 
        }

        \begin{table}[H]
            \adjustbox{max width=\textwidth, center = \textwidth}{%
            \begin{threeparttable}
            \caption{Employment Level Regressions}
            
            {
\def\sym#1{\ifmmode^{#1}\else\(^{#1}\)\fi}
\begin{tabular}{l*{6}{c}}
\toprule
                    &\multicolumn{2}{c}{\textbf{HQ Sample}}     &\multicolumn{2}{c}{\textbf{BR Sample}}     &\multicolumn{2}{c}{\textbf{Court Sample}}  \\
                    &\multicolumn{1}{c}{(1)}         &\multicolumn{1}{c}{(2)}         &\multicolumn{1}{c}{(3)}         &\multicolumn{1}{c}{(4)}         &\multicolumn{1}{c}{(5)}         &\multicolumn{1}{c}{(6)}         \\
\midrule
Bankruptcies        &     0.00596\sym{*}  &     0.00861\sym{**} &      0.0113\sym{**} &      0.0145\sym{***}&     0.00874\sym{**} &     0.00832\sym{**} \\
                    &   (0.00320)         &   (0.00375)         &   (0.00468)         &   (0.00542)         &   (0.00390)         &   (0.00410)         \\
\addlinespace
Forum Shopping      &                     &     -0.0127\sym{*}  &                     &     -0.0125         &                     &     0.00258         \\
                    &                     &   (0.00710)         &                     &   (0.00825)         &                     &   (0.00507)         \\
\midrule
County FE           &         Yes         &         Yes         &         Yes         &         Yes         &         Yes         &         Yes         \\
State-Year FE       &          No         &          No         &          No         &          No         &         Yes         &         Yes         \\
District-Year FE    &         Yes         &         Yes         &         Yes         &         Yes         &          No         &          No         \\
R-Squared           &       0.995         &       0.995         &       0.999         &       0.999         &       0.999         &       0.999         \\
Observations        &        5070         &        5070         &         354         &         354         &        1074         &        1074         \\
\bottomrule
\multicolumn{7}{l}{\footnotesize Standard errors clustered at county level in parentheses}\\
\multicolumn{7}{l}{\footnotesize \sym{*} \(p<0.1\), \sym{**} \(p<0.05\), \sym{***} \(p<0.01\)}\\
\end{tabular}
}

            \label{tab:RBST_HQ_BR_COURT}
            
            \begin{tablenotes}
                \item [] \footnotesize\textit{Note:} \tabnote{\lnLegEmp}{\regressorsNoShock}{\countyDisY or county and state-year}{\controls}{\clustercounty}{Columns (1) and (2) \sampleHQ; Columns (3) and (4) \sampleBRcounties; Columns (5) and (6) \sampleBRcourts}{\sampleyears}{\source}
            \end{tablenotes}
            \end{threeparttable}
            }
        \end{table} 
        
        \begin{table}[H]
            \adjustbox{max width=\textwidth, center = \textwidth}{%
            \begin{threeparttable}
            \caption{Employment Level Regressions}
            
            {
\def\sym#1{\ifmmode^{#1}\else\(^{#1}\)\fi}
\begin{tabular}{l*{6}{c}}
\toprule
                    &\multicolumn{1}{c}{(1)}         &\multicolumn{1}{c}{(2)}         &\multicolumn{1}{c}{(3)}         &\multicolumn{1}{c}{(4)}         &\multicolumn{1}{c}{(5)}         &\multicolumn{1}{c}{(6)}         \\
\midrule
Bankruptcies        &      0.0108\sym{***}&      0.0100\sym{**} &      0.0116\sym{**} &     0.00960\sym{**} &      0.0112\sym{**} &      0.0115\sym{***}\\
                    &   (0.00357)         &   (0.00394)         &   (0.00459)         &   (0.00448)         &   (0.00444)         &   (0.00402)         \\
\addlinespace
ln(County Population)&       1.104\sym{***}&       0.934\sym{***}&       0.892\sym{***}&       0.858\sym{***}&       0.917\sym{***}&       0.904\sym{***}\\
                    &    (0.0731)         &    (0.0743)         &    (0.0859)         &    (0.0868)         &    (0.0797)         &    (0.0785)         \\
\addlinespace
ln(Emp Non-Legal)   &       0.335\sym{***}&       0.221\sym{***}&       0.223\sym{***}&       0.230\sym{***}&       0.218\sym{***}&       0.224\sym{***}\\
                    &    (0.0387)         &    (0.0444)         &    (0.0471)         &    (0.0483)         &    (0.0458)         &    (0.0453)         \\
\midrule
                    &                     &                     &                     &                     &                     &                     \\
Fixed Effects       &                     &                     &                     &                     &                     &                     \\
County              &         Yes         &         Yes         &         Yes         &         Yes         &         Yes         &         Yes         \\
Year                &          No         &         Yes         &          No         &          No         &          No         &          No         \\
Region-Year         &          No         &          No         &         Yes         &          No         &          No         &          No         \\
Division-Year       &          No         &          No         &          No         &         Yes         &          No         &          No         \\
State-Year          &          No         &          No         &          No         &          No         &         Yes         &          No         \\
District-Year       &          No         &          No         &          No         &          No         &          No         &         Yes         \\
\midrule            &                     &                     &                     &                     &                     &                     \\
R-Squared           &       0.973         &       0.974         &       0.975         &       0.975         &       0.974         &       0.974         \\
Observations        &       29877         &       29877         &       29877         &       29822         &       29877         &       29877         \\
\bottomrule
\multicolumn{7}{l}{\footnotesize Standard errors clustered at county level in parentheses}\\
\multicolumn{7}{l}{\footnotesize \sym{*} \(p<0.1\), \sym{**} \(p<0.05\), \sym{***} \(p<0.01\)}\\
\end{tabular}
}

            \label{tab:RBST_90_02_BRonly}
            
            \begin{tablenotes}
                \item [] \footnotesize\textit{Note:} \tabnote{\lnLegEmp}{\regressorsBRonly}{\FEall}{\controls}{\clustercounty}{\samplerestr}{\sampleyearsExtendedPeriod}{\source}
            \end{tablenotes}
            \end{threeparttable}
            }
        \end{table}

    \subsection{Controls}
        In our benchmark specification, we include county level controls for the log of population and the log of the number of employees outside of the legal sector. In this subsection, Table \ref{tab:RBST_Lag_Unemp_Estab} extends our set of controls by lagging the controls and by including new time-varying county variables: unemployment rate, and percent change in number of non-legal establishments. 
        See Section \ref{sec:robust_check_controls} for a detailed discussion.

        \begin{table}[H]
            \adjustbox{max width=\textwidth, center = \textwidth}{%
            \begin{threeparttable}
            \caption{Employment Level Regressions}
            
            {
\def\sym#1{\ifmmode^{#1}\else\(^{#1}\)\fi}
\begin{tabular}{l*{6}{c}}
\toprule
                    &\multicolumn{2}{c}{\textbf{Lagged Controls}}&\multicolumn{2}{c}{\textbf{Unemployment Rate}}&\multicolumn{2}{c}{\textbf{\%$\Delta$(Non-Legal Estabs)}}\\
                    &\multicolumn{1}{c}{(1)}         &\multicolumn{1}{c}{(2)}         &\multicolumn{1}{c}{(3)}         &\multicolumn{1}{c}{(4)}         &\multicolumn{1}{c}{(5)}         &\multicolumn{1}{c}{(6)}         \\
\midrule
Bankruptcies        &     0.00803\sym{**} &      0.0101\sym{***}&     0.00758\sym{**} &     0.00998\sym{***}&     0.00794\sym{**} &      0.0104\sym{***}\\
                    &   (0.00329)         &   (0.00373)         &   (0.00334)         &   (0.00372)         &   (0.00348)         &   (0.00384)         \\
\addlinespace
Forum Shopping      &                     &    -0.00989         &                     &     -0.0114\sym{*}  &                     &     -0.0119\sym{*}  \\
                    &                     &   (0.00652)         &                     &   (0.00642)         &                     &   (0.00642)         \\
\addlinespace
ln(County Population)&                     &                     &       0.719\sym{***}&       0.719\sym{***}&       0.716\sym{***}&       0.716\sym{***}\\
                    &                     &                     &     (0.144)         &     (0.144)         &     (0.144)         &     (0.144)         \\
\addlinespace
ln(Emp Non-Legal)   &                     &                     &       0.175\sym{***}&       0.175\sym{***}&       0.161\sym{***}&       0.161\sym{***}\\
                    &                     &                     &    (0.0496)         &    (0.0496)         &    (0.0481)         &    (0.0481)         \\
\addlinespace
L.ln(County Population)&       0.855\sym{***}&       0.855\sym{***}&                     &                     &                     &                     \\
                    &     (0.149)         &     (0.149)         &                     &                     &                     &                     \\
\addlinespace
L.ln(Emp Non-Legal) &       0.126\sym{***}&       0.126\sym{***}&                     &                     &                     &                     \\
                    &    (0.0402)         &    (0.0402)         &                     &                     &                     &                     \\
\addlinespace
Unemployment Rate   &                     &                     &     0.00274         &     0.00275         &                     &                     \\
                    &                     &                     &   (0.00283)         &   (0.00283)         &                     &                     \\
\addlinespace
\%$\Delta$(Non-Legal Establishments)&                     &                     &                     &                     &      0.0774         &      0.0774         \\
                    &                     &                     &                     &                     &    (0.0722)         &    (0.0722)         \\
\midrule
County FE           &         Yes         &         Yes         &         Yes         &         Yes         &         Yes         &         Yes         \\
District-Year FE    &         Yes         &         Yes         &         Yes         &         Yes         &         Yes         &         Yes         \\
R-Squared           &       0.984         &       0.984         &       0.984         &       0.984         &       0.984         &       0.984         \\
Observations        &       15931         &       15931         &       15931         &       15931         &       15925         &       15925         \\
\bottomrule
\multicolumn{7}{l}{\footnotesize Standard errors clustered at county level in parentheses}\\
\multicolumn{7}{l}{\footnotesize \sym{*} \(p<0.1\), \sym{**} \(p<0.05\), \sym{***} \(p<0.01\)}\\
\end{tabular}
}

            \label{tab:RBST_Lag_Unemp_Estab}
            
            \begin{tablenotes}
                \item [] \footnotesize\textit{Note:} \tabnote{\lnLegEmp}{\regressorsNoShock}{\countyDisY}{Columns (1) and (2) \controlsLAG; Columns (3) and (4) \controlsUNEMPRATE; Columns (5) and (6) \controlsESTABCHANGE }{\clustercounty}{\samplerestr}{\sampleyears}{\sourceAVGWAGE}
            \end{tablenotes}
            \end{threeparttable}
            }
        \end{table}

    \subsection{Measurement}
        In this subsection, we explore the robustness of our main results to alternative measurements of treatment and employment. 
        The value weighted versions measure the total assets of all the bankrupt firms in billions of 2021 dollars.
        Table \ref{tab:RBST_Dummy_ValueWeight} gives alternative treatment measurement results by using binary indicators, or value weighted versions when measuring bankruptcies and forum shopping.
        Table \ref{tab:RBST_BLS_Adj} Columns (1) and (2) include an additional measure of the number of bankruptcies in adjacent counties to capture spillover effects; then
        Columns (3) and (4) use \BLS\space data instead of \CBP\space data to measure legal and non-legal employment for the dependent variable and controls respectively.
        In all cases, our findings are aligned with our priors.
        See Section \ref{sec:robust_check_measurement} for a detailed discussion. 
    
        {\setlength{\tabcolsep}{11pt}
        \begin{table}[H]
            \adjustbox{max width=\textwidth, center = \textwidth}{%
            \begin{threeparttable}
            \caption{Employment Level Regressions}
            
            {
\def\sym#1{\ifmmode^{#1}\else\(^{#1}\)\fi}
\begin{tabular}{l*{4}{c}}
\toprule
                    &\multicolumn{2}{c}{\textbf{Binary}}        &\multicolumn{2}{c}{\textbf{Value Weighted}}\\
                    &\multicolumn{1}{c}{(1)}         &\multicolumn{1}{c}{(2)}         &\multicolumn{1}{c}{(3)}         &\multicolumn{1}{c}{(4)}         \\
\midrule
Bankruptcy          &      0.0141         &      0.0174\sym{*}  &                     &                     \\
                    &   (0.00862)         &    (0.0101)         &                     &                     \\
\addlinespace
Forum Shop          &                     &     -0.0121         &                     &                     \\
                    &                     &    (0.0102)         &                     &                     \\
\addlinespace
BR Firm's Assets (2021 dollars)&                     &                     &     0.00219\sym{**} &     0.00281\sym{**} \\
                    &                     &                     &   (0.00103)         &   (0.00132)         \\
\addlinespace
FS Firm's Assets (2021 dollars)&                     &                     &                     &    -0.00198         \\
                    &                     &                     &                     &   (0.00204)         \\
\midrule
County FE           &         Yes         &         Yes         &         Yes         &         Yes         \\
District-Year FE    &         Yes         &         Yes         &         Yes         &         Yes         \\
R-Squared           &       0.984         &       0.984         &       0.984         &       0.984         \\
Observations        &       15931         &       15931         &       15931         &       15931         \\
\bottomrule
\multicolumn{5}{l}{\footnotesize Standard errors clustered at county level in parentheses}\\
\multicolumn{5}{l}{\footnotesize \sym{*} \(p<0.1\), \sym{**} \(p<0.05\), \sym{***} \(p<0.01\)}\\
\end{tabular}
}

            \label{tab:RBST_Dummy_ValueWeight}
            
            \begin{tablenotes}
                \item [] \footnotesize\textit{Note:} \tabnote{\lnLegEmp}{Columns (1) and (2) \regressorsDUMMY; Columns (3) and (4) \regressorsASSETS}{\countyDisY}{\controls}{\clustercounty}{\samplerestr}{\sampleyears}{\source}
            \end{tablenotes}
            \end{threeparttable}
            }
        \end{table}
        }

        {\setlength{\tabcolsep}{11pt}
        \begin{table}[H]
            \adjustbox{max width=\textwidth, center = \textwidth}{%
            \begin{threeparttable}
            \caption{Employment Level Regressions}
            
            {
\def\sym#1{\ifmmode^{#1}\else\(^{#1}\)\fi}
\begin{tabular}{l*{4}{c}}
\toprule
                    &\multicolumn{2}{c}{\textbf{Spillover}}     &\multicolumn{2}{c}{\textbf{BLS Data}}      \\
                    &\multicolumn{1}{c}{(1)}         &\multicolumn{1}{c}{(2)}         &\multicolumn{1}{c}{(3)}         &\multicolumn{1}{c}{(4)}         \\
\midrule
Bankruptcies        &     0.00658\sym{*}  &     0.00889\sym{**} &     0.00569\sym{**} &     0.00856\sym{***}\\
                    &   (0.00358)         &   (0.00399)         &   (0.00282)         &   (0.00321)         \\
\addlinespace
Forum Shopping      &                     &     -0.0109\sym{*}  &                     &     -0.0136\sym{***}\\
                    &                     &   (0.00633)         &                     &   (0.00496)         \\
\addlinespace
Adjacent County Bankruptcies&    -0.00337         &    -0.00336         &                     &                     \\
                    &   (0.00358)         &   (0.00358)         &                     &                     \\
\midrule
County FE           &         Yes         &         Yes         &         Yes         &         Yes         \\
District-Year FE    &         Yes         &         Yes         &         Yes         &         Yes         \\
R-Squared           &       0.984         &       0.984         &       0.996         &       0.996         \\
Observations        &       15931         &       15931         &       10684         &       10684         \\
\bottomrule
\multicolumn{5}{l}{\footnotesize Standard errors clustered at county level in parentheses}\\
\multicolumn{5}{l}{\footnotesize \sym{*} \(p<0.1\), \sym{**} \(p<0.05\), \sym{***} \(p<0.01\)}\\
\end{tabular}
}

            \label{tab:RBST_BLS_Adj}
            
            \begin{tablenotes}
                \item [] \footnotesize\textit{Note:} \tabnote{\lnLegEmp}{\regressorsADJ}{\countyDisY}{\controls}{\clustercounty}{\samplerestr}{\sampleyears}{\source}
            \end{tablenotes}
            \end{threeparttable}
            }
        \end{table}
        }

\section{GMM Estimation}\label{App:GMM}
This section describes our GMM estimation from Section \ref{sec:struct_est} in greater detail. We first describe the technical details of the estimation. Then we present some additional GMM robustness results to support our main findings in Table \ref{tab:gmm_results}. 





    \subsection{GMM Procedure}
    We implement this procedure using the Stata GMM package for more information please refer to the documentation for that package.
    For ease of notation let $i = 1,...,N$ index each observation where $N$ is the total number of county-year observations.  
    Let $\Zwi$ and $\Zni$ denote the vectors of instruments used for estimation. 
    The $1 \times q_w$ vector $\Zwi$ is used for $\Delta\ewsi$ and $\Delta\ewui$; the $1 \times q_n$ vector $\Zni$ is used for $\Delta \eSi$ and $\Delta \eUi$.
    Let $q = 2q_w + 2q_n$. 
    Then, for each observation, there are $q$ population moment conditions given by
    \[
    \E
    \begin{bmatrix}
    (\Zwi)'\cdot\Delta \ewsi  \\
    (\Zwi)'\cdot\Delta \ewui  \\
    (\Zni)'\cdot\Delta \eSi\\
    (\Zni)'\cdot\Delta \eUi
    \end{bmatrix}
    = \mathbf{0} \qquad \forall \; i = 1,...,N.
    \]
    For our preferred specification this corresponds to 14 moment conditions where
    \begin{gather*}
    \Zwi = 
    \begin{bmatrix}
    (\BR_{c,t-1}) & (\BR_{c,t-1})^2 & (BR_{c,t-1})^3 & (\BR_{c,t-1})^4
    \end{bmatrix},\\
    \Zni = \begin{bmatrix}
    \BR_{c,t} & (\BR_{c,t-1}) & 1
    \end{bmatrix}.
    \end{gather*}
    
    We denote the vector of parameters to be estimated as 
    $\be = \begin{bmatrix}
    1/\rho_s & 1/\rho_u & \alpha
    \end{bmatrix}'.$
    By stacking the moment equations we can write them more compactly as $\Zi'\ei(\be)$ with the $4\times q$ instrument matrix $\Zi$ and the $4\times1$ vector of residuals $\ei(\be)$ defined as
    \[
    \Zi \equiv 
    \begin{bmatrix}
    \Zwi & 0 & 0 & 0 \\
    0 & \Zwi & 0 & 0 \\
    0 & 0 & \Zni & 0 \\
    0 & 0 & 0 & \Zni
    \end{bmatrix}
    \qquad \text{and} \qquad
    \ei(\be) \equiv
    \begin{bmatrix}
    \Delta\ewsi \\
    \Delta\ewui \\
    \Delta\eSi \\
    \Delta\eUi \\
    \end{bmatrix}
    .\]
    Then the usual GMM criterion function can be written as 
    \begin{equation}
        Q(\be,\W) = \left\{\frac{1}{N}\sumin\Zi'\ei(\be)\right\}' \W \left\{\frac{1}{N}\sumin\Zi'\ei(\be)\right\}\;,
    \end{equation}
    for some positive semi definite $4\times 4$ weight matrix $\W$.
    
    The two-step GMM procedure first obtains a preliminary estimate of
        \[\hat{\be}_1=\argmin{\be} Q(\be,\hat{\W}_1),\] 
    using the identity matrix as the first stage weight matrix $\hat{\W}_1$. 
    We then update $\hat{\W}_2$ to be the estimate of the inverse variance-covariance matrix for $\Zi'\ei(\be)$ clustered at the county level.
    To do this first we calculate 
    \[
    \lc = \frac{1}{N}\sum_{i\in c}\Zi'\ei(\hat{\be}_1),
    \]
    for each county $c$. Then $\text{cov}(\Zi'\ei(\be))$ is estimated as 
    \[
    \boldsymbol{\Lambda} = \frac{1}{N}\sum_{c=1}^{N_c} \lc \lc'.
    \]
    where $N_c$ is the number of counties.
    Finally setting $\hat{\W}_2 =\boldsymbol{\Lambda}^{-1}$ for the second stage gives our GMM parameter estimates  
        \[\bgmm = \argmin{\be} Q(\be,\hat{\W}_2).\]
    
    Inference is done by calculating the usual variance covariance matrix
    \newcommand{\Gbar}{\bar{\mathbf{G}}}
    \[
    \text{Var}(\bgmm) = \frac{1}{N} \left(\Gbar(\bgmm)'\hat{\W}_2 \Gbar(\bgmm)\right)^{-1}
    \] 
    where
    \[
    \Gbar(\bgmm)=\frac{1}{N} \sumin \Zi' \;\frac{\partial\ei(\be)}{\be'}\Bigg|_{\be=\bgmm}. 
    \]

    \subsection{GMM Alternative Specifications}
        In order to address concerns regarding our GMM specification we present results for alternative specifications. 
        Table \ref{tab:gmm_appendix_inst_specs} details the instruments used for each equation in the alternative specifications, and Table \ref{tab:gmm_results_appendix} presents the results.
        Though many of the results are not as strong as our baseline specification, they all support our main findings presented in Section \ref{sec:GMM_results}.
    
        The first concern we will address regards our choice of sample. 
        We calibrate our preferred specification (Baseline) using 1991-2001 in order to get a larger sample size. 
        The \BLS\; data we use for skilled and unskilled has only 1208 observations when only the Court Competition Era is considered. 
        However when we limit the sample to only the Court Competition Era we find results that are very similar to our baseline specification albeit with lower significance due to the smaller sample size. 
        
        Next we address the choice of instruments in several ways. All specifications make use of the sources of exogenous variation in $\theta_{c,t}$ which are $\BR_{c,t}$ and $\BR_{c,t-1}$. 
        The first additional specification we check is the baseline specification without a constant term as an instrument for $\Delta \eSct$ and $\Delta \eUct$ to address concerns that $\E[\Delta \eSi]=0$ and $\E[\Delta \eUi] =0$ may not be valid moment conditions. 
        These results however support our main findings albeit with lower significance because when all instruments are constructed from $\BR_{c,t}$ and $\BR_{c,t-1}$ the model becomes harder to identify due to the fact that $\BR_{c,t}$ and $\BR_{c,t-1}$ equal zero for the majority of observations. 
        This brings us to the next additional specification where we include a constant term as an instrument for all of the moment conditions. 
        Including this specification gives the model considerably more variation to use for identification because the instruments $\BR_{c,t}$ and $\BR_{c,t-1}$ equal zero for the majority of observations. 
        With this additional variation we once again find results that support our baseline specification, and as expected has a J statistic that has a somewhat smaller $p$ value. 
        Finally to address concerns with including a variety of non-linear transformations of $\BR_{c,t}$ and $\BR_{c,t-1}$ as instruments we present a simplified instrument specification without these additional terms. 
        By not including these terms this greatly reduces the number of moment conditions and subsequently the amount of variation to be used for identification.
        As a result there is not enough variation in the model to properly identify the $\alpha$ returns to scale parameter, which gives an extremely large standard error for the estimate. However, even with larger standard errors, we find evidence to support our baseline models estimate of the elasticity of labor supply $1/\rho_s$ and $1/\rho_u$.        
        
        \begin{table}[H]
            \caption{Alternative GMM Instrument Specifications}
            \label{tab:gmm_appendix_inst_specs}
            \adjustbox{max width=1.02\textwidth, center = \textwidth}{%
            \centering
                \begin{threeparttable}
                    {
                    \begin{tabular}{l*{3}{c}}
                        \toprule
                        &\multicolumn{1}{c}{Instruments for $\Delta \epsilon_{c,t}^{wS}$ and $\Delta \epsilon_{c,t}^{wU}$}    
                        &\multicolumn{1}{c}{ Instruments for $\Delta \eSct$ and $\Delta \eUct$}    
                        \\
                        \midrule
                        Baseline
                            &  \{$(\BR_{c,t-1})$, $(\BR_{c,t-1})^2$, $(BR_{c,t-1})^3$, $(\BR_{c,t-1})^4$\}
                            & \{$\BR_{c,t}$, $(\BR_{c,t-1})$, constant\}
                            \\
                        \addlinespace
                        Court Competition Era   
                            &  \{$(\BR_{c,t-1})$, $(\BR_{c,t-1})^2$, $(BR_{c,t-1})^3$, $(\BR_{c,t-1})^4$\}
                            & \{$\BR_{c,t}$, $(\BR_{c,t-1})$, constant\}
                            \\
                        \addlinespace
                        No Constant
                            &  \{$(\BR_{c,t-1})$, $(\BR_{c,t-1})^2$, $(BR_{c,t-1})^3$, $(\BR_{c,t-1})^4$\}
                            & \{$\BR_{c,t}$, $(\BR_{c,t-1})$\}
                            \\
                        \addlinespace
                        With Constant
                            &  \{$(\BR_{c,t-1})$, $(\BR_{c,t-1})^2$, $(BR_{c,t-1})^3$, $(\BR_{c,t-1})^4$, constant\}
                            & \{$\BR_{c,t}$, $(\BR_{c,t-1})$, constant\}
                            \\
                            \addlinespace
                        Simplified Instruments 
                            &  \{$(\BR_{c,t-1})$\}
                            & \{$\BR_{c,t}$, constant\}
                            \\
                        \bottomrule
                    \end{tabular}
                    }
                \end{threeparttable}
            }
        \end{table}
    
        \begin{table}[H]
            \caption{Alternative GMM Instrument Specifications}
            \label{tab:gmm_results_appendix}
            \adjustbox{max width=1.02\textwidth, center = \textwidth}{%
                \centering
                \begin{threeparttable}
                    {
                    \begin{tabular}{l*{7}{c}}
                        \toprule
                        &\multicolumn{1}{c}{$1/\rho_s$}    
                        &\multicolumn{1}{c}{$1/\rho_u$}    
                        &\multicolumn{1}{c}{$\alpha$}
                        &\multicolumn{1}{c}{J-stat}
                        &\multicolumn{1}{c}{$EV_{s}$}
                        &\multicolumn{1}{c}{$EV_{u}$}
                        &\multicolumn{1}{c}{Obs.}
                        \\
                        
                        \midrule
                        Baseline    
                            & 0.0853\sym{***}
                            & 0.264\sym{**} 
                            & 1.292\sym{***} 
                            & 12.592
                            & 0.74\%
                            & 0.77\%
                            & 2370
                            \\
                            &    (0.0307)         
                            &     (0.120)         
                            &     (0.150)          
                            & ($p = 0.32$)
                            & & & 
                            \\
                        \addlinespace
                        Court Competition Era   
                            & 0.0792\sym{*}  
                            & 0.195         
                            & 1.397\sym{***}
                            & 16.921
                            & 0.20\%
                            & 0.21\%
                            & 1208
                            \\
                            & (0.0423)
                            & (0.164)
                            & (0.101)
                            & ($p = 0.11$)
                            & & &
                            \\
                        \addlinespace
                        No Constant
                            & 0.0223         
                            & 0.00933
                            & 1.268\sym{***}
                            & 10.2235
                            & 0.65\%
                            & 0.64\%
                            & 2370
                            \\
                            & (0.0519)
                            & (0.118) 
                            & (0.133)
                            & ($p = 0.33$)
                            & & &
                            \\
                        \addlinespace
                        With Constant
                            & 0.113\sym{***}
                            & 0.280\sym{**} 
                            & 1.294\sym{***}
                            & 17.8513
                            & 0.76\%
                            & 0.79\%
                            & 2370
                            \\
                            & (0.0289)
                            & (0.108) 
                            & (0.149)
                            & ($p = 0.16$)
                            & & &
                            \\
                            \addlinespace
                        Simplified Instruments 
                            & 0.0750\sym{**}
                            & 0.238\sym{*}
                            & -7.942
                            & 5.024
                            & 0.33\%
                            & 0.34\%
                            & 2370
                            \\
                            & (0.0325)
                            & (0.124)
                            & (18.16)
                            & ($p = 0.17$)
                            & & &
                            \\
                        \bottomrule
                        \multicolumn{5}{l}{\footnotesize Standard errors clustered at county level in parentheses}\\
                        \multicolumn{5}{l}{\footnotesize \sym{*} \(p<0.1\), \sym{**} \(p<0.05\), \sym{***} \(p<0.01\)}\\    
                    \end{tabular}
                    }
                \begin{tablenotes}
                    \item [] \footnotesize \textit{Note:} GMM regression estimates. Wages and unemployment are detrended prior to estimation using county-year population, log of county-year employment in all non-legal sectors with fixed effects for \countyDisY. \textit{Sample:} \sampleNYDE. \textit{Sources:} \sourceAVGWAGE. $EV_s$ $EV_u$ reported using $EIS = 0.5$ ($\sigma = 2$).  
                \end{tablenotes}
                \end{threeparttable}
            }
        \end{table}

\section{Database Creation}\label{App:Data_Creation}
    \begin{enumerate}
        \item Clean \Complong \space Database.%
            \footnote{Information on the S\&P Compustat data can be found in the  \href{http://sites.bu.edu/qm222projectcourse/files/2014/08/compustat\_users\_guide-2003.pdf}{Compustat User Guide}.}
        We do this in a similar manner to \citet{corbae_reorganization_2021}, for a more detailed description see Appendix A-1 in \citet{corbae_reorganization_2021}. 
            \begin{itemize}
                \item  Remove duplicate entries, remove non USA firms, remove firms in legal sector (SIC = 8111), and remove years outside of LP range (1980-2014). 
                \item Create bankruptcy indicators in a similar manner to \citet{corbae_reorganization_2021}. Ch11 if "fresh start accounting" and no deletion, Ch11 if "bankruptcy" and no deletion, Ch11 if "bankruptcy" and deletion with merger delete reason in CS.
            \end{itemize}
        
        \item Collect data for \Comp \space bankruptcies using court records.  
            \begin{itemize}
                \item Use bankruptcy indicators from \Comp \space to create list of companies that appear to have undergone Chapter 11 bankruptcy.  
                \item Use \SEC \space to find each companies EIN code.
                \item Search \PCLlong \space using companies EIN codes to find court records for bankruptcies. (Includes filing location) 
                \item Supplement missing cases with SEC filings found using \Bloomberg website when necessary.
            \end{itemize}
        
        \item Create firm-level database. 
            \begin{itemize}
                \item Merge \LP, \Comp, and data scraped from \PACER, to create complete panel data of publicly traded companies with information Chapter 11 bankruptcies. 
                \item Drop bankruptcy observations for legal industry from database. (SIC = 8111)
                \item Determine bankruptcy occurred when indicated by either \LP \space or one of the bankruptcy indicators from \Comp.
                \item Determine if forum shopping occurred if either indicated by \LP \space or if the \PACER \space filing location does not match the district where the firm was headquartered (Note this does NOT capture within district forum shopping!). If filing location data missing, assume firm did not forum shop.   
            \end{itemize}
            
        \item Create County-Year panel data used for regressions.
            \begin{itemize}
                \item Aggregate firm-level database to county-year level and merge with \CBP \space employment data. 
                \item Include \Census, \BLS, and \BEA \space data for controls and additional exercises. 
            \end{itemize}
    \end{enumerate}

\pagebreak

\end{document}